\documentclass[11pt,preprint]{aastex}
\pdfoutput=1
\usepackage{bm}
\usepackage{graphicx}
\usepackage{amsmath}    % Advanced maths commands
\usepackage{amssymb}    % Extra maths symbols
\usepackage{footnote}
\usepackage{color}
\usepackage{amsmath}
\usepackage{mathtools}
\usepackage{pdflscape}
\usepackage{multirow}

\let\oldfootsep=\footnotesep
\newcommand\ltsima{$\; \buildrel <\over\sim \;$}
\newcommand\simlt{\lower.5ex\hbox{\ltsima}}
\newcommand\gtsima{$\; \buildrel >\over\sim \;$}
\newcommand\simgt{\lower.5ex\hbox{\gtsima}}

\newcommand\msun {M_\odot}

\newcommand{\mathbold}[1]{\mbox{\boldmath $\bf#1$}}
\newcommand\piEbold{{\mathbold \pi_E}}
\newcommand\mubold{{\mathbold \mu}}

\newcommand\vbold{{\mathbold v}}
\shorttitle{}
\shortauthors{Bennett et al}

\begin{document}


\title{Keck Observations Confirm a Super-Jupiter Planet Orbiting M-dwarf OGLE-2005-BLG-071L}


\author{David~P.~Bennett\altaffilmark{1,2},
Aparna~Bhattacharya\altaffilmark{1,2},
Jean-Philippe~Beaulieu$^{3,4}$,
Joshua~W.~Blackman\altaffilmark{3},
Aikaterini~Vandorou\altaffilmark{3},
%Ian~A.~Bond\altaffilmark{5},
Sean K.~Terry\altaffilmark{5},
Andrew A.~Cole\altaffilmark{3},
Calen B.~Henderson\altaffilmark{6},
Naoki~Koshimoto$^{1,2,7}$,
Jessica~R.~Lu\altaffilmark{8},
Jean~Baptiste~Marquette\altaffilmark{9}, 
 Cl{\'e}ment~Ranc\altaffilmark{1},
Andrzej~Udalski$^{10}$
 \\  } 
              
\keywords{gravitational lensing: micro, planetary systems}

\affil{$^{1}$Code 667, NASA Goddard Space Flight Center, Greenbelt, MD 20771, USA;    \\ Email: {\tt bennettd@umd.edu}}
\affil{$^{2}$Department of Astronomy, University of Maryland, College Park, MD 20742, USA}
\affil{$^{3}$School of Physical Sciences, University of Tasmania, Private Bag 37 Hobart, Tasmania 7001 Australia}
\affil{$^{4}$Institut d'Astrophysique de Paris, 98 bis bd Arago, 75014 Paris, France}
%\affil{$^{5}$Institute of Natural and Mathematical Sciences, Massey University, Auckland 0745, New Zealand}
\affil{$^{5}$Department of Physics, Catholic University of America, 620 Michigan Ave., N.E. Washington, DC 20064, USA}
%\affil{$^{5}$Institute of Space and Astronautical Science, Japan Aerospace Exploration Agency, Kanagawa 252-5210, Japan}
%\affil{$^{8}$Department of Physics, University of Auckland, Private Bag 92019, Auckland, New Zealand}
%\affil{$^{11}$Department of Earth and Space Science, Graduate School of Science, Osaka University, Toyonaka, Osaka 560-0043, Japan}
\affil{$^{6}$NASA Exoplanet Science Institute, IPAC/Caltech, Pasadena, California 91125, USA}
\affil{$^{7}$Department of Astronomy, Graduate School of Science, The University of Tokyo, 7-3-1 Hongo, Bunkyo-ku, Tokyo 113-0033, Japan}
\affil{$^{8}$University of California Berkeley, Berkeley, CA}
\affil{$^{9}$Laboratoire d'astrophysique de Bordeaux, Univ. Bordeaux, CNRS, B18N, allée Geoffroy Saint-Hilaire, 33615 Pessac, France}
\affil{$^{10}$Warsaw University Observatory, Al.~Ujazdowskie~4, 00-478~Warszawa,Poland}


\begin{abstract}
We present adaptive optics imaging from the NIRC2 instrument on the Keck-2 telescope that 
resolves the exoplanet host (and lens) star as it separates from the brighter source star. These observations 
yield the $K$-band brightness of the lens and planetary host star, as well as the lens-source relative proper motion,
$\mubold_{\rm rel,H}$. in the heliocentric reference frame. The $\mubold_{\rm rel,H}$ measurement 
allows determination of the microlensing parallax vector, $\piEbold$, which had only a single component
determined by the microlensing light curve. The combined measurements of $\mubold_{\rm rel,H}$ and $K_L$
provide the masses of the host stat, $M_{\rm host} = 0.426\pm 0.037 \msun$, and planet, 
$m_p = 3.27 \pm 0.32 M_{\rm Jupiter}$ with a projected separation of  $3.4\pm 0.5\,$AU. 
This confirms the tentative conclusion of a previous paper \citep{dong-ogle71} that this super-Jupiter
mass planet, OGLE-2005-BLG-071Lb, orbits an M-dwarf. Such planets are predicted to be rare by
the core accretion theory and have been difficult to find with other methods, but there are two
two such planets with firm mass measurements from microlensing, and an additional 11 planetary microlens
events with host mass estimates $< 0.5\msun$ and planet mass estimates $> 2$ Jupiter masses that could
be confirmed by high angular follow-up observations. We also point out that OGLE-2005-BLG-071L has 
separated far enough from its host star that it should be possible to measure the host star metallicity with
spectra from a high angular resolution telescope such as Keck, the VLT, the Hubble Space Telescope or the James
Webb Space Telescope.
\end{abstract}

%\clearpage

\section{Introduction}
\label{sec-intro}
The second exoplanet found by the microlensing method was OGLE-2005-BLG-071Lb, but the
discovery paper did not do a detailed analysis of the higher order microlensing effects that could
constrain the masses and distance of the lens system \citep{udalski05}. A more detailed
analysis was done in a follow-up paper \citep{dong-ogle71} that included high angular resolution 
Hubble Space Telescope (HST) observations. Unfortunately, the light curve had no significant 
signal for finite source effects, the HST observations were taken too soon after the event to get a
strong measurement of the lens-source relative proper motion, $\mubold_{\rm rel}$, and the light
curve only constrains one component of the two-dimensional microlensing parallax vector,
$\piEbold$. However, the light curve measurements of $\piEbold$, $\mubold_{\rm rel}$, were 
marginally inconsistent with each other, and the lack of light curve finite source effects prevented
an additional estimate of the angular Einstein radius. Furthermore, \citet{dong-ogle71} pointed
out that their estimates of the lens system masses and distance were dependent on the assumption
that the excess flux seen at the location of the source in the HST images was due to the lens star,
rather than a companion to the source. Nevertheless, with the assumption that the excess flux was
due to the lens and planetary host star, they conclude that OGLE-2005-BLG-071Lb may be the 
most massive planet known to orbit an M-dwarf at the time of its publication in 2009.

In this paper, we use adaptive optics (AO) observations with the NIRC2 instrument on the Keck-2 telescope
to identify the lens and planetary host star and provide a precise measurement of the masses and
distance of the OGLE-2005-BLG-071L planetary system. These new results largely confirm the 
more tentative conclusions of \citet{dong-ogle71}, except for the conclusion that the lens system has
thick disk kinematics. 

Despite the decade between the publication of \citet{dong-ogle71} and this paper, there are still very
few known massive planets in wide orbits around low-mass stars. NASA's exoplanet 
archive\footnote{https://exoplanetarchive.ipac.caltech.edu/} lists only 19 planets with masses $> 2M_{\rm Jupiter}$ 
with semi-major between the snow line (taken to be at $2.7 (M/\msun)\,$AU) and $30\,$AU
orbiting stars main sequence stars
with $0.08 \msun \leq M \leq 0.5\msun$. However, two of these are actually planets found by radial 
velocities with apparent typographical errors in the host masses, so that the correct masses are above $0.5\msun$.
The only remaining planet found by radial velocities in this category is GJ 317 b \citep{johnson07},
which has host and planet masses of $M_{\rm host} = 0.42 \pm 0.05\msun$ and 
$m_p = 2.5^{+0.7}_{-0.4}M_{\rm Jupiter}$ with a semi-major axis of $a =1.15 \pm 0.05\,$AU, based
on a combination of radial velocity and astrometric data \citep{anglada-escude12_gj317}. This implies 
an overlap between the planet's semi-major axis and our estimated snow line position of
$d_{\rm snow} = 2.7 (M/\msun)\,{\rm AU} = 1.161\pm 0.135\,$AU \citep{kennedy_snowline}.

The remaining 16 of these wide orbit planets with super-Jupiter mass planets listed in
the NASA Exoplanet Archive have been found by microlensing. These may seem to challenge the
core accretion theory expectation that super-Jupiter mass planets should be rare in orbits around
low-mass stars \citep{laughlin04}, but we must be careful to distinguish between planet and host star
mass measurements and Bayesian mass estimates based on the assumption that all stars are equally
likely to host a planet of the measured mass ratio, $q$. Recent AO image analysis for events MOA-2007-BLG-400
(A. Bhattacharya et al., in preparation) and MOA-2012-BLG-220 \citep{van19}
has indicated host masses much larger than the previous Bayesian analyses suggested, indicating that 
this prior assumption that all stars are equally likely to host gas giant planets may be flawed. 
Therefore, it is important to focus on events with mass measurements.

The first microlens exoplanet with a confirmed planet mass of $> 2M_{\rm Jupiter}$ orbiting a low-mass star with 
$0.08 \msun \leq M \leq 0.5\msun$ was OGLE-2012-BLG-0406 \citep{poleski_ob120406}, which has host 
and planet masses of $M_{\rm host} = 0.44 \pm 0.07\msun$ and $m_p = 2.73 \pm 0.43 M_{\rm Jupiter}$ 
according to a detailed analysis by \citet{tsapras_ob120406}. In this paper, we present the second such
microlens planet, OGLE-2005-BLG-071Lb with host 
and planet masses of $M_{\rm host} = 0.431 \pm 0.034\msun$ and $m_p = 3.37 \pm 0.30 M_{\rm Jupiter}$ 
at a projected separation of $3.4 \pm 0.5\,$AU. However, event OGLE-2005-BLG-071Lb is included in the
combined statistical sample of \citet{suzuki16}, which means that it can readily be included in
a statistical analysis of planet properties. In fact, there are three other events from the  \citet{suzuki16}
sample that should allow host and planet mass measurements from high angular resolution follow-up
observations. These are OGLE-2008-BLG-355 \citep{koshimoto14}, MOA-2009-BLG-387 \citep{batista11} 
and MOA-2011-BLG-322Lb \citep{shvartzvald14}, while such measurements for MOA-2012-BLG-006 
\citep{poleski_mb12006} in the \citet{suzuki16} sample would be quite difficult, because of its bright
giant source star. In addition to these events in the \citet{suzuki16} sample, the following events 
are candidates for super-Jupiter planets orbiting low-mass ($< 0.5 \msun$) stars:
MOA-2010-BLG-73,
OGLE-2013-BLG-0102, OGLE-2013-BLG-1761, OGLE-2015-BLG-0954, MOA-2016-BLG-227,
OGLE-2016-BLG-0263, KMT-2016-BLG-1397, and KMT-2017-BLG-1038
\citep{street_mb10073,jung_ob130102,hirao17,shin16,bennett_ob150954,kosh17_mb16227,zang18_kb161397,shin19_kb171038_1146}. 
Perhaps the most interesting candidate is OGLE-2018-BLG-1011 \citep{han_ob181011}, which has two 
super-Jupiter planets
with mass ratios of $0.015$ and $0.0095$ orbiting a star with an estimated mass of $\sim 0.2\msun$. 

This paper is organized as follows. In Section~\ref{sec-Keck}, we describe the Keck high angular 
resolution follow-up observations and their analysis, and in Section~\ref{sec-murel}, we discuss
the measured lens-source relative proper motion and how it can be used to constrain the microlensing
parallax and angular Einstein radius. Our preliminary light curve modeling is presented in 
Section~\ref{sec-lc_model}. In Section~\ref{sec-parallax}, we explain how we constrain the 
light curve modeling in order to sample the light curves that are consistent with the data, and then we 
present the lens system masses, separation and distance that are implied by the combined light curve
and Keck follow-up data in Section~\ref{sec-lens_prop}. Finally, we discuss the implications of these
results and present our conclusions in Section~\ref{sec-conclude}.

\section{Keck Follow up Observations and Analysis}
\label{sec-Keck}

\begin{figure}
\epsscale{0.9}
\plotone{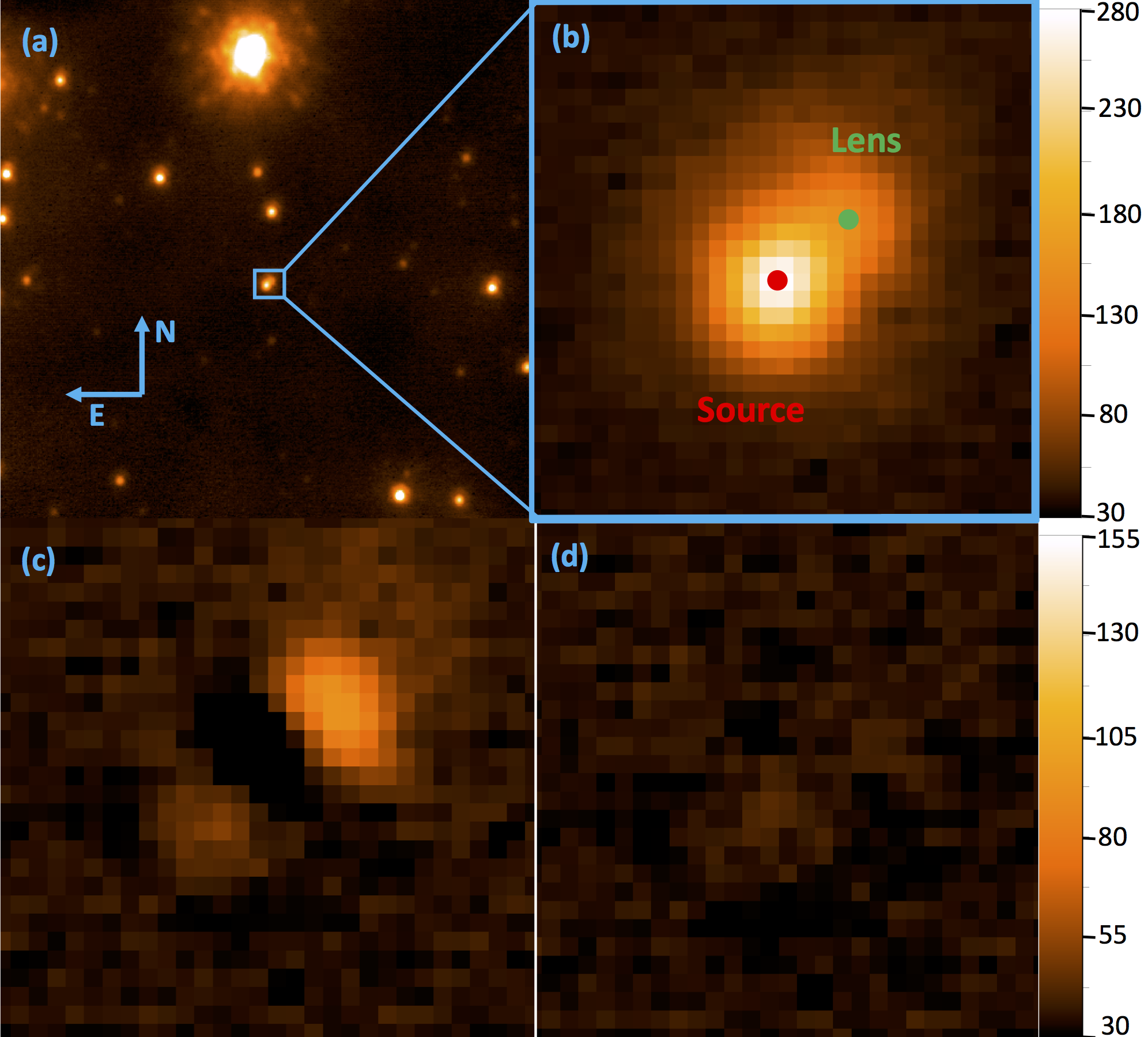}
\caption{(a) The coadded sum of 23 60-sec exposure Keck-NIRC2 narrow camera images, and (b) a closeup of the
OGLE-2005-BLG-071 source and lens stars on the right. The lens star is located $41.5\pm 1.3\,$mas West 
and $36.1\pm 1.3\,$mas North of the source star. (c) is the residual image from a DAOPHOT single star fit,
and (d) is the residual image from a two-star fit. The color-bar on the top right refers to panels (a) and (b), and the
color bar on the bottom right refers to both panels (c) and (d).
\label{fig-Keck}}
\end{figure}

We observed the source and lens stars for microlensing event OGLE-2005-BLG-071 with the
NIRC2 instrument on the Keck-2 telescope on May 28, 2019 as a part of our NASA Keck Key 
Strategic Mission Support program entitled ``Development of the 
Wide Field Infrared Survey Telescope (WFIRST) Exoplanet Mass Measurement Method\rlap".
The observations were carried out using both the NIRC2 wide and narrow cameras, which employ 
$1024\times 1024$ pixels with image scales of 39.686\,mas/pixel and 9.942\,mas/pixel, respectively.
All Keck images were taken using the Keck-2 laser guide star AO system.
The wide images are used for photometric calibrations, and we used a co-add of 8 of 10 dithered, 
wide camera images in the $K_S$ passband to calibrate to reprocessed images from the
VVV survey \citep{minniti-vvv}, which is calibrated to 
2MASS \citep{2mass_cal} following \citet{beaulieu18}. 
(Two of these wide camera images were not used due to relatively poor image quality.)
These wide camera images were flat field and dark current 
corrected using standard methods. We use  Scamp \citep{Scamp2010} to perform astrometry 
on the individual frames. We then did a median stacking using the SWarp package \citep{SWarp,Swarp2010}.
 
The details of our methods are described in \citet{batista14}. We  performed  aperture photometry on these wide 
camera images using  the latest version of the SExtractor code \citep{sextractor}. These wide images were used 
to detect and match as many bright isolated stars as possible to our custom reduction \citep{beaulieu16} of the 
calibrated VVV images. Once the wide camera images were calibrated, we calibrated the narrow 
camera images to the wide camera images. This procedure provided an overall calibration precision for
the co-added narrow camera image based on nominal uncertainties of 0.06 magnitudes. 
SExtractor, Scamp and SWarp are distributed by astromatic.net \citep{Astromatic2012}.

%For the reduction of these images, we used $K$-band dome flats 
%taken with narrow camera on the same day as the science images. There were 5 dome flat  images with 
%the lamp on and 5 more images with the lamp off, each with 65 seconds exposure time. Also at the end of 
%the night, we took 20 sky images using a clear patch of sky at a (RA, Dec) of (18:08:04.62, -29:43:53.7) 
%with an exposure time  of 30 seconds each. All these images were used to flat field, bias subtract and 
%remove bad pixels and cosmic rays from the raw science images. Finally 
%these clean raw images were stacked into one image and that we used for the final photometry and
%astrometry analysis. 

We took 40 dithered NIRC2 narrow camera were taken on the same night as the wide camera images,
and the 23 best images were combined to make the image
shown in Figures~\ref{fig-Keck}(a) and (b) after correcting for 
achromatic differential refraction \citep{refraction} and geometric distortion \citep{distortion}. 
Chromatic differential refraction is negligible compared to our measurement uncertainties \citep{gubler},
so we ignore it.
The point spread function (PSF) full-width, half-max (FWHM) of this image is $\approx 65\,$mas.
Because the source and candidate lens stars are separated by $\sim 1\,$FWHM, we must analyze the Keck
data with a PSF fitting code to measure the astrometry and photometry of this 2-star system. Following
\citet{aparna18}, we use DAOPHOT \citep{Daophot} for this analysis, which has a proven ability to handle 
peculiar PSF shapes that are sometimes encountered in AO images \citep{bennett-ogle109}.
The initial DAOPHOT reduction did not detect the fainter component of the source plus lens star blend 
because
it does not try to find stars at such small separations. This
resulted in the residual image shown in Figure~\ref{fig-Keck}(c) when the best fit PSF model was subtracted from 
the the original image (shown in Figures~\ref{fig-Keck}(a) and (b)). Following
\citet{aparna18}, we then added an additional source at the
position of the brightest residual in Figure~\ref{fig-Keck}(c) and reran DAOPHOT. This resulted in the two
stars indicated in Figure~\ref{fig-Keck}(b) and the two-star fit residual shown in Figure~\ref{fig-Keck}(d).
This two-star fit residual is nearly featureless, indicating that the Keck image is well modeled by 
two stars separated by $\sim 1\,$FWHM.
This analysis reveals that the magnitudes of the two stars are $K_S = 17.679 \pm 0.060$ and 
$K_L = 18.925 \pm 0.062$, where the uncertainty is dominated by the calibration uncertainty. 
The identification of the brighter star as the source star is determined
from the predicted $K$-band brightness of the source star.

In order to determine which star is the source and which is the candidate lens star, we need to compare
to the light curve predictions. The \citet{dong-ogle71} paper gives magnitudes and colors based on an early 
version of the OGLE-III survey calibrations and uses old values for the intrinsic color and magnitude of
red clump giant stars to estimate the extinction. We redo this analysis with updated OGLE-III light curves,
with the final instrumental magnitudes, the final OGLE-III magnitude and color calibrations 
\citep{ogle3-phot} and the \citet{nataf13} values for the properties of the bulge red clump giant
stars at the Galactic position ($l = -4.4198^\circ$, $b = -3.7864^\circ$) of the target. Using standard
methods \citep{bennett-ogle109}, we find the centroid of the red clump for stars within $2^\prime$ of the
source to be $[(V-I)_{\rm rcg}, I_{\rm rcg}]= [1.85, 15.71]$. Based on the intrinsic red clump giant stars
from \citet{nataf13}, this gives a color excess of $E(V-I) = 0.79$ and $I$-band extinction of $A_I = 1.10$,
which together imply $V$-band extinction of $A_V = 1.89$. The $K$-band extinction was 
determined to be $A_K = 0.15\pm 0.05$ from the online VVV extinction calculator \citep{vvv_extinct} using the 
\citet{nish09} extinction law. From the best fit constrained light curve model, described in Section~\ref{sec-parallax},
we find a source color of $V_S-I_S = 1.43$, so the extinction corrected color is $V_{S0}-I_{S0} = 0.64$.
Using the color-color relations of \citet{kenyon95} and the $I$-band magnitude, $I_S=19.54$ from this same model, 
we predict a source magnitude of $K_S = 17.89$. This can only be consistent with the brighter of the
two blended stars shown in Figures~\ref{fig-Keck}(a) and (b). The fit source brightness is fainter than the this
brighter star by about 3-$\sigma$,
which suggests that the source might have a faint companion. 

\section{The Lens-Source Relative Proper Motion}
\label{sec-murel}

The measured offset of the lens star with respect to the source is $\Delta{\rm RA} = -41.5\pm 1.3\,$mas and
$\Delta{\rm DEC} = 36.1\pm 1.3\,$mas. The Keck observations were taken on 28 May 2019, 
which was 14.103 years after the event. The apparent motion relative motion of the lens and source
is primarily due to their space motions, but there is also a contribution from parallax due to the
orbital motion of the Earth. However, this contribution is $\simlt 0.2\,$mas for a lens more
distant that $D_L \simgt 2\,$kpc. We can ignore this contribution, because it is 
much smaller than the error bars on our 
position measurements, so the lens-source relative proper motion is
\begin{equation}
\mubold_{\rm rel,H} = (\mu_{{\rm rel,H,E}}, \mu_{{\rm rel,H,N}})  = (-2.945 \pm 0.091, 2.563 \pm 0.091)\,{\rm mas/yr}\ ,
\label{eq-mu_rel_H}
\end{equation}
where the 
subscript H indicates that this is the proper motion in the Heliocentric reference frame, while
the E and N subscripts indicate the East and North directions. We can also convert these relative proper
motions to Galactic coordinates: 
\begin{equation}
\mubold_{\rm rel,H} = (\mu_{{\rm rel,H},l}, \mu_{{\rm rel,H},b})  = (0.703\pm 0.091, 3.840\pm 0.091)\,{\rm mas/yr}\ .
\label{eq-mu_rel_Hgal}
\end{equation}
The uncertainties in the separation measurements were calculated following \citet{king1983}, using the
equation
\begin{equation}
\sigma_{x} = 0.65238 \times \rm{FWHM} \times \sqrt{\frac{4}{3}}\times \frac{\sigma_{ \rm{F}}}{\rm {F}} 
\end{equation}
where F is the flux of the star with the measured position. 

\subsection{Source and Lens Proper Motion}
\label{sec_mu}

In addition to the lens-source relative proper motion, we can also compare the astrometry of
our 2019 Keck images to the astrometry of Hubble Space Telescope (HST) images taken
in 2005 and 2006 \citep{dong-ogle71}. We select the 2005 $V$-band (F555W) images taken 14.013 years
before the Keck images because these will provide a negligible parallax signal while providing
a precise position for source star, since the the lens-source separation is quite small one month
after the event peak, and the source was still magnified by a factor of $\sim 2$ and is much
brighter than the lens in the $V$-band. The HST data was reduced with DOLPHOT \citep{hstphot}
which does not produce astrometry with precision that is quite as high as the method of
\citet{andking00,andking04,andking06}, but this will have little influence on the final proper motion
results, because the uncertainties are dominated by the uncertainties in the Gaia astrometry.

There are three relatively bright stars within $5^{\prime\prime}$ of the OGLE-2005-BLG-071 
microlensing event with proper motion measurements in the Gaia DR2 data release \citep{gaia_dr2}. These stars
have Gaia numbers 4041487538966873216, 4041487538966873344, and 4041487538966886016, and
all three of these events have reliable proper motion measurements as evidenced by their
renormalized unit weight error (RUWE) values of $< 1.3$ \citep{lindegren18}.
Their $V$-band magnitudes are 18.399, 18.238, and 16.953, respectively. This allows us to determine the 
source proper motions in Galactic coordinates:
\begin{equation}
\mubold_{S} = (\mu_{S,l}, \mu_{S,b})  = (-4.99 \pm 0.19, -0.69 \pm 0.25)\,{\rm mas/yr}\ .
\label{eq-mu_S_gal}
\end{equation}
We can then add $\mubold_{S}$ and $\mubold_{\rm rel,H} $ from equation~\ref{eq-mu_rel_Hgal} to
determine the lens proper motion, 
\begin{equation}
\mubold_{L} = (\mu_{L,l}, \mu_{L,b})  = (-3.14 \pm 0.21, 3.38 \pm 0.27)\,{\rm mas/yr}\ .
\label{eq-mu_L_gal}
\end{equation}

The source proper motion, $\mubold_{S}$, is dominated by the reflex motion of
Galactic rotation and is quite typical for the proper motion of Galactic bulge star.
\citet{dong-ogle71} claimed that the OGLE-2005-BLG-071L lens star had the kinematics 
of a thick disk star. If true, this would be a surprise because super-Jupiter mass planets are
thought to form preferentially around more massive, high metallicity stars. Thick disk stars
tend to have low metallicity, so it would be surprising if OGLE-2005-BLG-071Lb was a
super-Jupiter mass planet orbiting a a low-mass, low metallicity star. 
The \citet{dong-ogle71} claim of thick disk kinematics is based upon their 
claim of velocity with respect to the local standard of rest of $v_{\rm LSR} = 103\pm 15\,$km/sec.
We can compare this to our measurement of the lens star transverse velocity if we assume
a lens distance $D_L = 3.5\,$kpc, which is the favored $D_L$ according to the analysis presented
below in Section~\ref{sec-lens_prop}. We find a transverse velocity of 
$\vbold_L  = (v_{L,ll}, v_{L,b}) = (-52.5, 56.5)\,$km/sec for a total transverse velocity of $77\,$km/sec.
This is within 2-$\sigma$ of the \citet{dong-ogle71} $v_{\rm LSR}$, but in fact, \citet{dong-ogle71}
do not specify that their $v_{\rm LSR}$ value refers to a transverse velocity, so it is probably 
supposed to be a full three-dimensional velocity, with the unmeasured radial component just an
estimate from averaging over a Galactic model. This would imply that their $v_{\rm LSR}$ value
is probably consistent with our measured $77\,$km/sec transverse velocity. However, their 
conclusion that the OGLE-2005-BLG-071L has thick disk kinematics does not follow.

The Gaia DR2 data release enables a much more accurate test of the kinematic properties 
for stars that are a few kpc from the Sun. In particular, the \citet{gaia18} study of the Milky Way
disk kinematics provides valuable information on the kinematic properties of stars in the vicinity of
OGLE-2005-BLG-071L., located at a distance of $D_L \sim 3.5\,$kpc toward the Galactic bulge
at $\sim 0.22\,$kpc below the Galactic plane. Figure 11 of \citet{gaia18} indicates that the median
orbital velocity in the disk has dropped to $v_\phi = 205\,$km/sec compared to $v_\phi = 240\,$km/sec at the 
Solar circle. This means that the velocity of OGLE-2005-BLG-071L compared to the stars in
its vicinity $\vbold_{L,{\rm VSR}}  = (-17.5, 56.5)\,$km/sec (where VSR stands for vicinity standard
of rest). The median stellar velocity in the $z$ direction (perpendicular to the Galactic disk) doesn't
differ from the median velocity of 0 at the Solar circle, but the velocity dispersions in both directions
grow to $\sigma_{v_\phi} = 45\,$km/sec and $\sigma_{v_z} = 38\,$km/sec. Thus, the $l$ and $b$
components of $\vbold_L$ are 0.4-$\sigma$ and 1.5-$\sigma$ from the center of the velocity 
distribution. So, we conclude that OGLE-2005-BLG-071L has normal stellar kinematics for 
a star at its location, and it should not be considered to be a thick disk star.

\subsection{Geocentric Relative Proper Motion}
\label{sec-murelG}

Our light curve modeling is done in a geocentric reference frame that differs from the heliocentric
frame by the instantaneous velocity of the Earth at the time of peak magnification. This avoids 
large corrected uncertainties for the light curve parameters that would be quite common for
models done in the heliocentric frame. However, this also means that
the lens-source relative proper motion that we measure with follow-up observations is not in the
same reference frame as the light curve parameters. This is an important issue because, as we
show below, the measured relative proper motion can be combined with the microlensing parallax 
light curve parameter to determine the mass of the lens system. The relation between the relative
proper motions in the heliocentric and geocentric coordinate systems are given by \citep{dong-ogle71}:   
\begin{equation}
\bm{\mu}_{\rm rel,H} = \bm{\mu}_{\rm rel,G} + \frac{{\bm v}_{\oplus} \pi_{\rm rel}}{\rm AU}  \ ,
\label{eq-mu_helio}
\end{equation}
where ${\bm v}_{\oplus}$ is the projected velocity of the earth relative to the sun (perpendicular to the 
line-of-sight) at the time of peak magnification. The projected velocity for OGLE-2005-BLG-071 is
${{\bm v}_{\oplus}}_{\rm E, N}$ = (15.508, 4.685) km/sec = (3.271, 0.988) AU/yr at the peak of the 
microlensing light curve, HJD'= 3480.70. 
The relative parallax is 
defined as $\pi_{\rm rel} \equiv 1/D_L - 1/D_S$, where $D_L$ and $D_S$ are lens and source 
distances, so equation~\ref{eq-mu_helio} can be written as:
\begin{align*} 
\bm{\mu}_{\rm rel,G} = \bm{\mu}_{\rm rel,H} - (3.271, 0.988)\times (1/D_L - 1/D_S) \ ,
\end{align*}
which is a more convenient form since $\bm{\mu}_{\rm rel,H}$ has been measured directly from the Keck images.
Now at each possible lens distance, 
we can use the $\mu_{\rm rel,G}$ value from equation~\ref{eq-mu_helio} to determine the angular 
Einstein radius, $\theta_E = \mu_{\rm rel,G} t_E$. As we explain below, the $\bm{\mu}_{\rm rel,G}$
can also be used to convert a one-dimensional microlensing parallax measurement into a full
measurement of the microlensing parallax vector. The lens flux and $\bm{\mu}_{\rm rel,H}$ measurements 
from the Keck observations 
and the 1-D parallax measurement constrain the angular Einstein radius, 
the microlensing parallax vector and, therefore, the mass and distance of the lens.

\section{Light Curve Model}
\label{sec-lc_model}

 \citet{dong-ogle71} present a detailed discussion of light curve models for OGLE-2005-BLG-071, and
we have little to add to this discussion. However, we find it necessary to do our own modeling of this event
in order to apply the constraints from our Keck observations in a Markov Chain Monte Carlo (MCMC) analysis.
We use the same data set as  \citet{dong-ogle71} with some minor modifications. First, we drop the 
Palomar data set because it covers only 80 minutes when the magnification changes by less than 5\%. 
Because the source and background fluxes are fit independently for each data set
\citep{rhie_98smc1}, the Palomar data do 
not provide a significant constraint on the light curve model. As is common practice
\citep{dong-ogle71}, once we have established
all the possible degenerate solutions, we renormalize the  $\chi^2$ per degree of freedom to be
$\chi^2/{\rm d.o.f.}\simeq 1$ in order to obtain more reasonable error bars on the inferred parameters
of the lens system.

\begin{deluxetable}{cccccc}
\tablecaption{Best Fit Model Parameters Unconstrained by Keck or HST Observations
                         \label{tab-UCmparams} }
\tablewidth{0pt}
\tablehead{
%% Use a footnote to explain numbering.
%& & & & \multicolumn{2}{c} {MCMC averages} \\
& \multicolumn{2}{c} {$u_0 < 0$} & \multicolumn{2}{c} {$u_0 > 0$} &  \\
\colhead{parameter}  & \colhead{$s<1$} & \colhead{$s> 1$} & \colhead{$s<1$} & \colhead{$s>1$} &\colhead{MCMC averages}
}  % end header.
\startdata
%                      plpce_3   plpfe_3     plpoe_3   plpne_3  
$t_E$ (days) & 77.100 & 70.942 & 72.324 & 69.825 & $71.8\pm 3.2$  \\   
$t_0$ (${\rm HJD}^\prime$) & 3480.7048 & 3480.7048 & 3480.6971 & 3480.6971 & $3480.702\pm 0.004$  \\
$u_0$ & -0.021476 & -0.027294 & 0.027680 & 0.023016 & $0.006\pm 0.025$  \\
$s$ & 0.75384 & 1.29784 & 0.75805 & 1.29140 & $1.15 \pm 0.24$  \\
$\alpha$ (rad) & -1.65212 & -1.64712 & 1.65041 & 1.64547 & $-0.40\pm 1.60$  \\
$q \times 10^{3}$ & 6.5824 & 7.1689 & 6.8499 & 7.1594 & $6.98 \pm 0.29$  \\
$t_\ast$ (days) & 0.01026 & 0.01652 & 0.02106 & 0.03092 & $0.018\pm 0.012$ \\
$\pi_{\rm E,N}$ & -0.9150 & -0.6331 & 0.08204 & -0.1065 & $-0.42\pm 0.41$ \\
$\pi_{\rm E,E}$ & -0.3692 & -0.3058 & -0.21894 & -0.2311 & $-0.279\pm 0.066$\\
fit $\chi^2$ & 1241.91 & 1226.24 & 1250.50 & 1231.14 &  \\
dof & 1271 & 1271 & 1271 & 1271 \\
\enddata
\end{deluxetable}

Table~\ref{tab-UCmparams} shows the parameters of our four degenerate light curve models and
the Markov Chain average of all four models. The parameters that apply to single lens models are
the Einstein radius crossing time, $t_E$, the time of closest alignment between the source and the
lens system center-of-mass, $t_0$, and the distance of closest approach between the source and the lens 
system center-of-mass, $u_0$, which is given in units of the Einstein radius. The addition of a second lens mass
requires three additional parameters, the mass ratio of the two lens masses, $q$, their separation, $s$, in
units of the Einstein radius, and angle, $\alpha$, between the source trajectory and the transverse 
line that passes through the two lens masses. In addition, a large fraction of binary lens systems exhibit
finite source effects that can be modeled with the addition of the source radius crossing time parameter, $t_*$.
For high magnification events, like OGLE-2005-BLG-071, the transformation $s \rightarrow 1/s$ often 
has only a slight change on the shape of the light curve. This is known as the close-wide degeneracy,
and it applies to OGLE-2005-BLG-071 \citep{udalski05,dong-ogle71}.

Many longer duration binary lens events, like OGLE-2005-BLG-071, exhibit light curve effects due to the
orbital motion of the Earth during the event. This is known as the microlensing parallax effect, and 
\citet{dong-ogle71} have shown that OGLE-2005-BLG-071 has a significant signal, which is described by
a two-dimensional vector, $\piEbold$, that is parallel to the direction of lens-source relative motion.
The inclusion of the microlensing parallax parameters introduces an addition degeneracy. Without
microlensing parallax, the transformation $u_0 \rightarrow -u_0$, $\alpha \rightarrow -\alpha$ produces 
exactly the same light curve, and it can just be thought of as a change of parameters. When microlensing
parallax is added, however, this transformation amounts to a reflection of the lens system orientation with
respect to the orbit of the Earth, so it is no longer a trivial, exact degeneracy (unless the lens system
lies in the ecliptic plane). This second degeneracy is often referred to as the $u_0 \leftrightarrow -u_0$
degeneracy. We do not include lens orbital motion for these unconstrained models, which we consider only to 
offer a comparison to the 
models with models with the constraints from high angular resolution follow-up observations.

There is one fixed parameter not included in Table~\ref{tab-UCmparams}. The geocentric 
coordinate system used to define the microlensing parallax parameters is fixed to the 
Earth's orbital velocity at $t_{\rm fix} = 3480.7$ These unconstrained models have 1298 observations, 
9 non-linear parameters, and 18 linear parameters for a total of 1271 degrees of freedom.

\section{Relative Proper Motion Constraints on ${\bm{\pi}_{\rm E}}$ and Light Curve Models}
\label{sec-parallax}

\begin{figure}
\epsscale{0.9}
\plotone{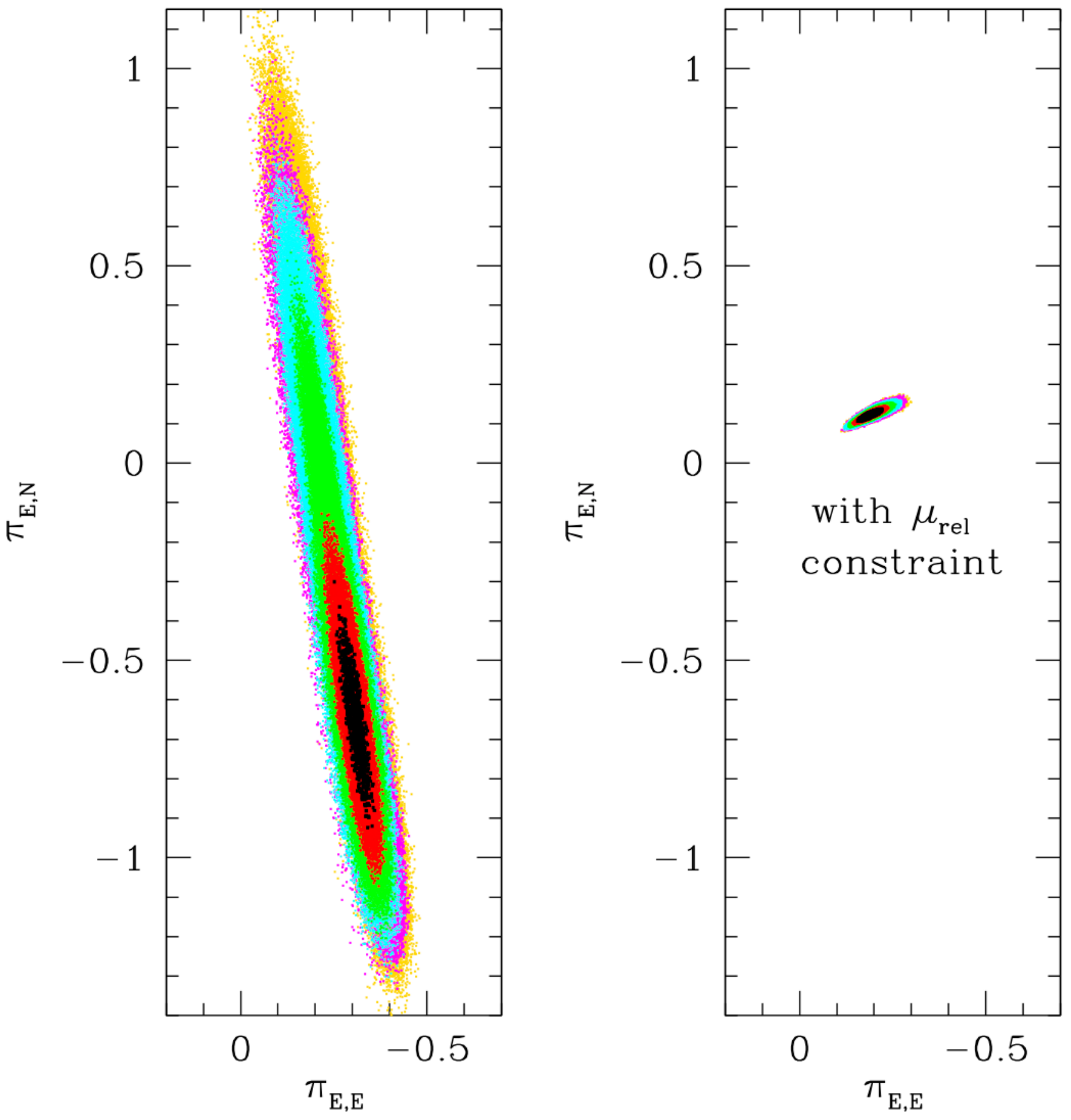}
\caption{{\it Left panel}: The ${\bm \pi_{\rm E}}$ distribution from light curve modeling without any constraints
from follow-up observations.
{\it Right panel}: The ${\bm \pi_{\rm E}}$ distribution resulting from the addition of the high resolution follow-up
imaging constraints. The following color scheme is used to denote the $\chi^2$ differences from the best fit 
light curve model: black represents $\Delta\chi^2 < 1$, red represents $\Delta\chi^2 < 4$, 
green represents $\Delta\chi^2 < 16$, cyan represents $\Delta\chi^2 < 25$, and magenta represents 
$\Delta\chi^2 \geq 25$. (The weighting of the models does not depend on the $\Delta\chi^2$ values,
however.) The right panel shows that the Keck relative proper motion measurements 
constrain the North component of ${\bm \pi_{\rm E}}$ ($\pi_{\rm E, N}$), which
was largely unconstrained by the light curve. Without the $\bm{\mu}_{\rm rel,H}$ measurement, in the left
panel, the light curve slightly favors solutions with $\pi_{\rm E, N} < 0$, but the constraint forces 
$\pi_{\rm E, N} > 0$. Note that this figure combines both the degenerate $u_0 > 0$ and $u_0 < 0$ models.}
\label{fig-parallax}
\end{figure}

As Figure~\ref{fig-parallax} indicates, the microlensing light curve provides a fairly tight constraint
on one component of the microlensing parallax vector, $\piEbold$. This is the component in the 
direction of the Earth's acceleration, which is nearly parallel to the East-West direction.
But as it is often the case
\citep{muraki11}, only the $\pi_{E,E}$ 
component of the microlensing parallax vector is measured precisely. As shown in the left panel of 
Figure~\ref{fig-parallax}, the 2-$\sigma$ range for $\pi_{E,N}$ is $-0.39 < \pi_{E,N} < 0.43$. However,
the microlensing parallax vector, ${\bm{\pi}_{\rm E}}$ is parallel to the $\bm{\mu}_{\rm rel,G}$ vector, and 
the two quantities are related by
\begin{equation}
{\bm{\pi}_{\rm E}} = \frac{\pi_{\rm rel}}{t_{\rm E}}\frac{\bm{\mu}_{\rm rel,G}}{|\mu_{\rm rel,G}|^2} \ ,
\label{eq-piE_muG}
\end{equation}
so with measurements of $\pi_{E,E}$ and $\bm{\mu}_{\rm rel,H}$, we can use equations~\ref{eq-mu_helio}
and \ref{eq-piE_muG} to solve for $\pi_{E,N}$ \citep{gould-diskorhalo,ghosh04,bennett07}. 
While this leads to a quadratic equation
\citep{gould-1dpar}, only one of these solutions has a positive lens distance, $D_L$.
\citet{aparna18} have argued that the odds of an ambiguity due the two solutions of this 
quadratic equation are negligible, because even is cases where both solutions have positive $D_L$,
they predict very different lens magnitudes.

The volume of the light curve model parameter space that is allowed by the Keck observation constraints on 
$\bm{\mu}_{\rm rel,H}$ and $K_L$ is very much smaller than the volume allowed without those constraints
as Figure~\ref{fig-parallax} indicates. As a result, we would have poor sampling of the posterior distributions
if we were to apply these constraints to Markov Chains. So, following \citet{aparna18} we implement constraints
on $\bm{\mu}_{\rm rel,H}$ and $K_L$ during the light curve modeling. The relative proper motion, $\mubold_{\rm rel,H}$, 
is constrained by the measurement given in equation~\ref{eq-mu_rel_H}, and we constrain the lens
star $K$-band brightness to be $K_L = 18.925 \pm 0.062$, as discussed in Section~\ref{sec-Keck}.
Following \citet{aparna18}, these constraints implement a $\chi^2$ contribution from each of the 
constraints and adding it to the light curve fit $\chi^2$ inside the modeling code \citep{bennett-himag}.
We use the empirical mass-luminosity relation described by \citet{bennett_moa291}, which is a combination of
several different mass-luminosity relations for different mass ranges. 
For $M_L \geq 0.66\,\msun$, $0.54\,\msun\geq M_L \geq 0.12\,\msun $, and 
$0.10 \,\msun \geq M_L \geq 0.07\,\msun$, we use the relations of \citet{henry93}, \citet{delfosse00},
and \citet{henry99}, respectively. In between these
mass ranges, we linearly interpolate between the two relations used on the
boundaries. That is, we interpolate between the \citet{henry93} and the \citet{delfosse00}
relations for $0.66\,\msun > M_L > 0.54\,\msun$, and we interpolate between the
\citet{delfosse00} and \citet{henry99} relations for $0.12\,\msun > M_L > 0.10\,\msun$.

For the mass-luminosity relations, we must also consider the foreground extinction.
At a Galactic latitude of $ b = -3.7865^\circ$, and a lens distance of $\sim 2\,$kpc, the lens system
is likely to be behind some, but not all, of the dust that is in the foreground of the source. 
We assume a dust scale height of $h_{\rm dust} = 0.10\pm 0.02\,$kpc, so that the
extinction in the foreground of the lens is given by
\begin{equation}
A_{i,L} = {1-e^{-|D_L(\sin b)/h_{\rm dust}|}\over 1-e^{-|D_S (\sin b)/h_{\rm dust}|}} A_{i,S} \ ,
\label{eq-A_L}
\end{equation}
where the index $i$ refers to the passband: $I$, $V$, or $K$. In the Markov Chain calculations themselves,
we fix $D_S = 8.8\,$kpc for our source star at a Galactic longitude of $l = -4.4198^\circ$, and we fix the
dust scale height at $h_{\rm dust} = 0.10\,$kpc. (The distance to Galactic bar at this longitude is $8.8\,$kpc
\citep{nataf13}.)
But, we remove this distance restriction by reweighting the
links in the Markov Chain when we sum them for our final results as discussed in Section~\ref{sec-lens_prop}.

\begin{deluxetable}{cccccc}
\tablecaption{Best Fit Model Parameters with $\mubold_{\rm rel,H}$ and Magnitude Constraints
                         \label{tab-Cmparams} }
\tablewidth{0pt}
\tablehead{
%% Use a footnote to explain numbering.
%& & & & \multicolumn{2}{c} {MCMC averages} \\
& \multicolumn{2}{c} {$u_0 < 0$} & \multicolumn{2}{c} {$u_0 > 0$} &  \\
\colhead{parameter}  & \colhead{$s<1$} & \colhead{$s> 1$} & \colhead{$s<1$} & \colhead{$s> 1$} &\colhead{MCMC averages}
}  % end header.
\startdata
%                      plmuopae_9   plmuopde_9   plmuopoe_9   plmuopne_9 
$t_E$ (days) & 68.357 & 68.221 & 68.420 & 67.912 & $68.1\pm 1.2$  \\   
$t_0$ (${\rm HJD}^\prime$) & 3480.6751 & 3480.6919 & 3480.6755 & 3480.6922 & $3480.692\pm 0.007$  \\
$u_0$ & -0.024468 & -0.028289 & 0.024401 & 0.028439 & $0.001\pm 0.028$  \\
$s_0$ & 0.76151 & 1.28821 & 0.76067 & 1.28956 & $1.233 \pm 0.163$  \\
$\alpha$ (rad) & -1.64182 & -1.64233 & 1.64218 & 1.64267 & $0.06\pm 1.64$  \\
$q \times 10^{3}$ & 7.0943 & 7.2753 & 7.1262 & 7.3421 & $7.32 \pm 0.16$  \\
$t_\ast$ (days) & 0.05103 & 0.05113 & 0.05087 & 0.05143 & $0.0513\pm 0.0014$ \\
$\pi_{\rm E,N}$ & 0.1223 & 0.1234 & 0.1218 & 0.1204 & $0.123\pm 0.010$ \\
$\pi_{\rm E,E}$ & -0.1899 & -0.1913 & -0.1685 & -0.1909 & $-0.192\pm 0.021$\\
$\dot{s}_x$(days$^{-1}$) & 0.00139 & -0.00126 & 0.00139 & -0.00128 & $-0.00081\pm 0.00084$ \\
$\dot{s}_y$(days$^{-1}$) & -0.00187 & -0.00128 & 0.00173 & 0.00102 & $-0.00002\pm 0.00081$ \\
$I_s$ & 19.550 & 19.543 & 19.554 & 19.539 & $19.542\pm 0.026$  \\
$V_s$ & 20.981 & 20.975 & 20.971 & 20.986 & $20.974\pm 0.026$  \\
fit $\chi^2$ & 1234.29 & 1230.13 & 1234.27 & 1229.98 &  \\
dof & $\sim 1273$ & $\sim 1273$ & $\sim 1273$ & $\sim 1273$ \\
\enddata
\end{deluxetable}

\begin{figure}
\epsscale{0.9}
\plotone{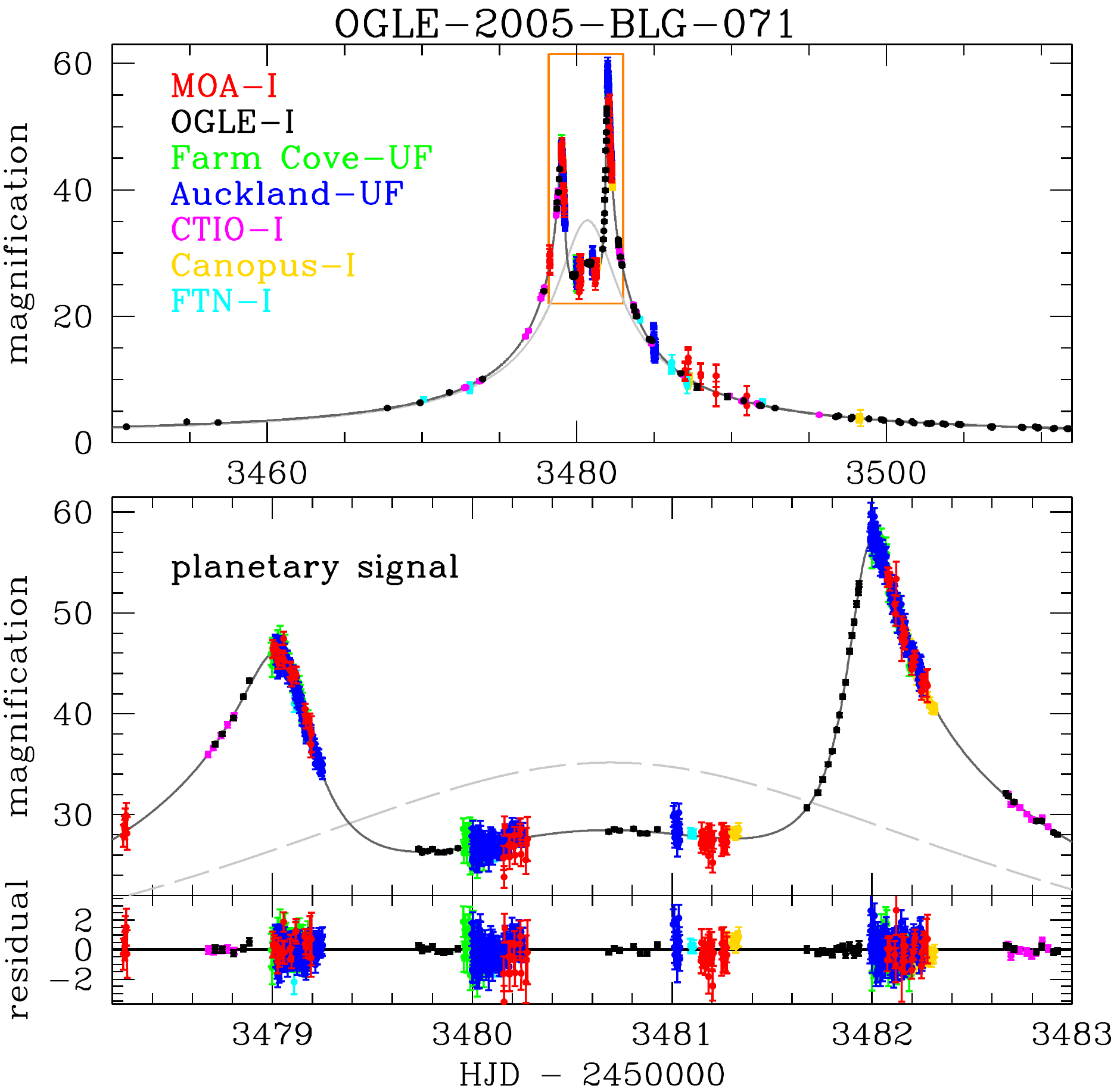}
\caption{The best fit light curve with the constraints on the relative proper motion, $\mubold_{\rm rel,H}$,
and host star magnitude, $K_L$ constraints. This is the model from the fifth column of 
Table~\ref{tab-Cmparams}, with $u_0 > 0$ and $s > 1$.}
\label{fig-lc}
\end{figure}

In addition to these constraints, we also include two additional parameters to describe the orbital motion 
of the planet. Unlike the case of the two planet event OGLE-2006-BLG-109 \citep{gaudi-ogle109,bennett-ogle109},
the light curve does not provide significant constraints on the planetary orbital motion, 
although we fix the orbital period at 5000 days to avoid the possibility of an unphysical future
encounter between the source and lens system that might occur if we assumed constant velocities. 
The two orbital motion parameters
are $\dot{s}_x$ and $\dot{s}_y$, which are the orbital motion in the direction of star-planet separation 
(at $t = t_{\rm fix}$) and the perpendicular direction, respectively. As in the case
of MOA-2009-BLG-266 \citep{muraki11}, the inclusion of orbital motion does affect the relative weighting of the
degenerate models. In particular, the inclusion of orbital motion substantially decreases the $\chi^2$ 
difference between the wide and close models, which are only disfavored by $\Delta\chi^2 \approx 4.3$
for the constrained models. 
This can be seen by a comparison of the $\chi^2$ difference between the wide and close models for
the constrained model, shown in Table~\ref{tab-Cmparams} and the unconstrained models without
orbital motion, shown in Table~\ref{tab-UCmparams}. As was the case with MOA-2009-BLG-266, the 
$\chi^2$ difference between the $u_0 >0$ and $u_0 < 0$ solutions is also decreased. The best fit 
constrained light curve is shown in Figure~\ref{fig-lc}. Despite the two additional light curve parameters,
the model constraints on the relative proper motion and lens magnitude increase the number of degrees
of freedom to 1272 or 1274, depending on whether the upper limits on the combined lens plus source
$V$ and $I$ magnitudes are counted as full constraints.

While these constraints have little effect on the best fit model $\chi^2$, they have a dramatic
effect on the allowed range of microlensing parallax parameters, as Figure~\ref{fig-parallax}
indicates. The 2-$\sigma$ range for $\pi_{E,N}$ is reduced from $-1.07 < \pi_{E,N} < 0.49$
to $0.104 < \pi_{E,N} < 0.144$, which is a reduction of a factor of 39 in uncertainty. This yields a microlensing parallax
amplitude of $\pi_E = 0.229 \pm 0.023$, which will be used in Section~\ref{sec-lens_prop} to determine the 
lens mass.

The orbital motion parameters in these models are constrained to be ensure consistency
with a bound orbit, using the method presented by \citet{muraki11}. The last column of Table~\ref{tab-Cmparams}
provides the mean and RMS of each of the model parameters, and distributions of these parameters
are approximately symmetric.

\section{Lens Properties}
\label{sec-lens_prop}

The situation for OGLE-2005-BLG-071 is similar to the case of OGLE-2012-BLG-0950 \citep{aparna18}
in that the source radius crossing time, $t_*$, has not been reliably measured. Fortunately, as discussed 
above in Section~\ref{sec-murelG}, it is straight forward to constrain the angular Einstein radius, $\theta_E$,
with our measurement of $\bm{\mu}_{\rm rel,H}$ given in equations~\ref{eq-mu_rel_H} and \ref{eq-mu_rel_Hgal}.
The conversion between the heliocentric and geocentric frames only provides a minor complication. 
This simplest if we assume that the 
source distance, $D_S$, is known. In this case, we can determine $\bm{\pi}_{\rm E}$ 
and $\bm{\mu}_{\rm rel,G}$ from the measurement
of $\bm{\mu}_{\rm rel,H}$ and light curve model parameters using equations \ref{eq-mu_helio} and \ref{eq-piE_muG}.
We can then determine
the angular Einstein radius from $\theta_E = t_E \mu_{\rm rel,G}$. 
 
\begin{figure}
\epsscale{1.0}
\plotone{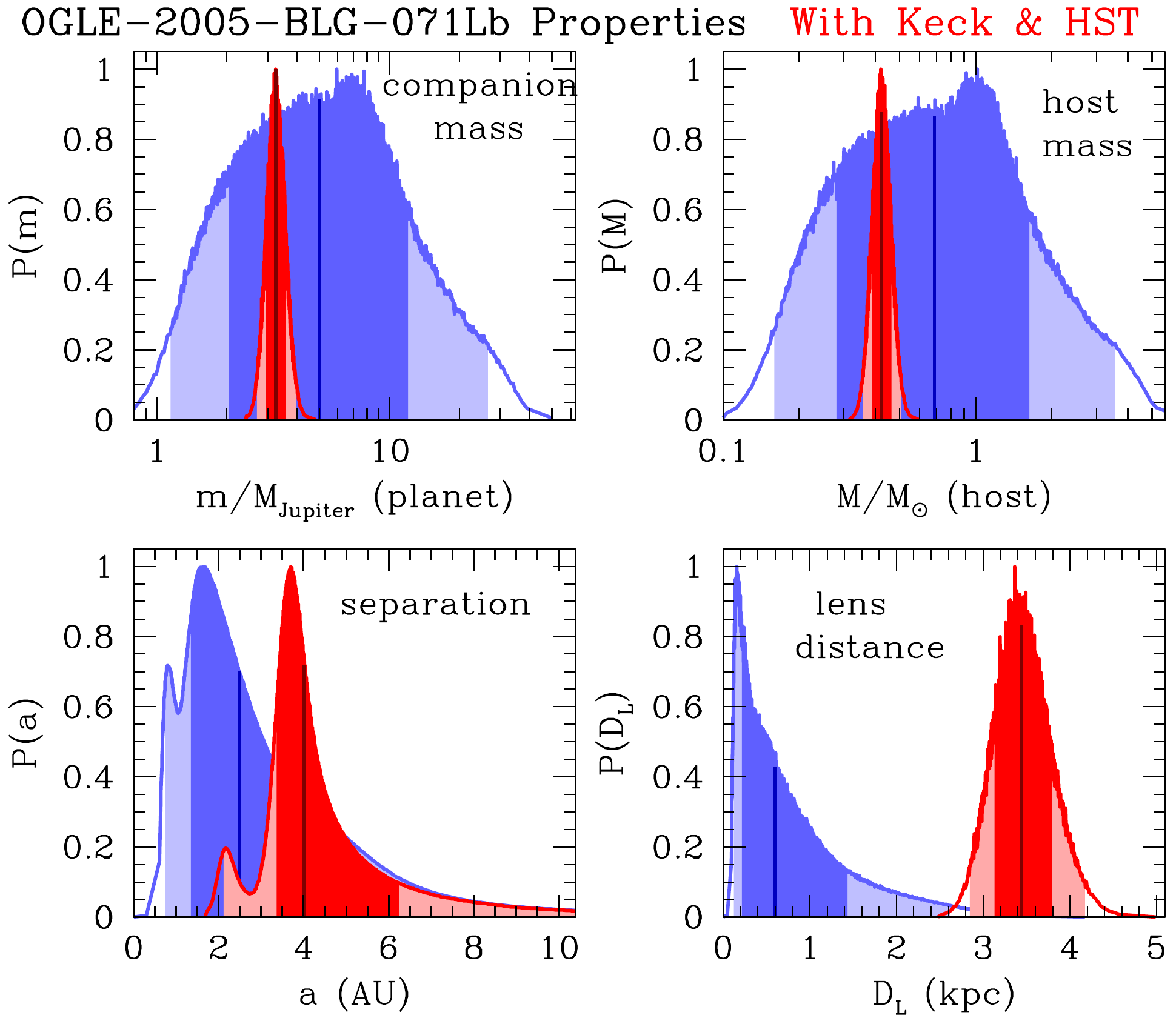}
\caption{The Bayesian posterior probability distributions for the planetary companion mass, host mass, 
their separation and the distance to the lens system are shown with only light curve constraints in blue 
and with the additional constraints from our Keck and HST follow-up observations in red.
The central 68.3$\%$ of the distributions are shaded in darker colors (dark red and dark blue) and the 
remaining central 95.4$\%$ of the distributions are shaded in lighter colors. The vertical black line marks 
the median of the probability distribution of the respective parameters.}
\label{fig-lens_prop}
\end{figure}

Measurements of either angular Einstein radius, $\theta_E$, or the microlensing parallax 
amplitude, $\pi_E$, provide mass-distance relations \citep{bennett_rev,gaudi_araa},
\begin{equation}
M_L = {c^2\over 4G} \theta_E^2 {D_S D_L\over D_S - D_L} 
       =  {c^2\over 4G}{ {\rm AU}\over{\pi_E}^2}{D_S - D_L\over D_S  D_L}  \ ,
\label{eq-m_thetaE}
\end{equation}
which can be combined to yield the lens mass in an expression with no dependence on the lens or source distance,
\begin{equation}
M_L = {c^2 \theta_E {\rm AU}\over 4G \pi_E} = {\theta_E \over (8.1439\,{\rm mas})\pi_E} \msun \ .
\label{eq-m}
\end{equation}
The lens system distance can also be determined from
\begin{equation}
D_L={{\rm AU}\over \pi_{\rm E}\theta_E+1/D_S} \ ,
\label{eq-Dl}
\end{equation}
but it does depend on $D_S$.
Our measurement of the host star $K$-band magnitude, $K_L$, also implies a mass-distance relation
when combined with our empirical mass-luminosity relation, and the HST observation in 2005 and 2006
\citep{dong-ogle71} provide upper limits on the host star brightness in the $I$ and $V$ bands.
As a result, this system is overconstrained with mass measurements from 
equation~\ref{eq-m} and the combination of equation~\ref{eq-Dl} and the $K$-band mass-luminosity relation. 
In our MCMC light curve models, we have fixed $D_S$, and
applied $\chi^2$ constraints based on the measured $\bm{\mu}_{\rm rel,H}$ and $K$-band values and the
$I_L$ and $V_L$ upper limits.

We follow a somewhat similar procedure in our constrained light curve modeling. We include both 
components of $\piEbold$ and $t_*$ as model parameters even though the light curve data provide
very weak constraints on $\pi_{\rm E,N}$ and $t_*$. With the $\pi_E$ and $t_*$ values from each light
curve model, we can determine $\theta_E$ and then $M_L$ from equation~\ref{eq-m}. With $D_S$ fixed,
we use equation~\ref{eq-Dl} to determine $D_L$. This provides all the information that we need to determine
$K_L$, $I_L$, and $V_L$ from the mass-luminosity relations and equation~\ref{eq-A_L} and $\bm{\mu}_{\rm rel,G}$
from equation~\ref{eq-mu_helio}. We then calculate addition contributions to $\chi^2$ for each of these 
five measured parameters (two components of $\bm{\mu}_{\rm rel,H}$ and three magnitudes). These 
$\chi^2$ contributions have the form $\Delta\chi^2 = e^{-(y_{\rm model}-y_{\rm meas})^2/2\sigma_y^2}$
where $y_{\rm meas}$ and $\sigma_y$ measured value and uncertainty, while $y_{\rm model}$ is the
model value. The magnitudes are converted to linear fluxes before the constraints are applied, and the 
$V$ and $I$ constraints are only applied if the model flux is larger than the measured values.

For our final results, we combine the parallel Markov chains for our light curve models, but we reweight
each of the models to remove the fixed distance constraint. We multiply by the inverse of the weighting
with the fixed source distance applied in the light curve model, and then we select a new $D_S$ value from
the probability distribution from the Galactic model from \citet{bennett14}.

 \begin{deluxetable}{cccc}
\tablecaption{Measurement of Planetary System Parameters from the Lens Flux Constraints\label{tab-params}}
\tablewidth{0pt}
\tablehead{\colhead{parameter}&\colhead{units}&\colhead{values \& RMS}&\colhead{2-$\sigma$ range}}
\startdata
Angular Einstein Radius, $\theta_E$&mas&$0.793\pm 0.042$&0.710--0.876 \\
Geocentric lens-source relative proper motion, $\mu_{\rm rel, G}$&mas/yr&$4.25\pm  0.21$&3.84--4.66\\
Host star mass, $M_{\rm host}$&${\msun}$&$0.426\pm 0.037$ & 0.357--0.506\\
Planet mass, $m_p$&$M_{\rm Jup}$& $3.27\pm 0.32$& 2.70--3.96\\
Host star - Planet 2D separation, $a_{\perp}$&AU&$3.38\pm 0.52$& 1.95--4.08\\
Host star - Planet 3D separation, $a_{3\rm d}$&AU&$4.0^{+2.3}_{-0.6}$& 2.1--14.3\\
Lens distance, $D_L$&kpc &$3.46\pm 0.33$& 2.85--4.18\\
Source distance, $D_S$ &kpc & $9.28\pm 1.17$& 6.70--11.14\\
\enddata
\end{deluxetable}

Figure~\ref{fig-lens_prop} and Table~\ref{tab-params} provide the results of our analysis. We find that the
host star has a mass of $M_{\rm host} = 0.43\pm 0.04\msun$ and it is orbited by a super-Jupiter mass 
planet with $m_p = 3.3\pm 0.3 M_{\rm Jup}$ at a projected separation of $a_\perp = 3.4\pm 0.5\,$AU.
This translates to a three-dimensional separation of $a_{3\rm d} = 4.0^{+2.3}_{-0.6}\,$AU under the
assumption of a random orientation of the planetary orbit, and the lens
system is located at a distance of $D_L = 3.5\pm 0.3\,$kpc. These distributions are indicated by the 
red histograms in Figure~\ref{fig-lens_prop}.
These results are a dramatic improvement in precision over blue histograms that indicate the 
parameters predicted by our Bayesian analysis without any constraints from Keck or HST observations. 
However, the lens distance distribution predicted by the unconstrained analysis is peculiar with its sharp
peak at very small distances. This is due to the slight light curve preference for very small source radius
crossing time, $t_*$, values. We suspect that this is due to systematic errors in the Auckland Observatory
photometry that our overcome by the much stronger constraints on $\bm{\mu}_{\rm rel,H}$ and $K_L$
from the Keck images. As discussed by \citet{dong-ogle71}, the flat field images from this telescope had serious 
scattered light problems that were exacerbated by a $180^\circ$ flip of the telescope orientation in the 
middle of each night. Also, the observations were unfiltered, which led to photometry errors as large
as 7\% due to differential atmospheric chromatic extinction. It is possible that some of these systematic errors remain
after \citet{dong-ogle71} attempts to correct them. For example, their correction for differential chromatic extinction
implicitly assumed that the atmospheric chromatic extinction effects remained constant over all four days
when the event was at its peak, but the water vapor and aerosol composition of the atmosphere might
have changed during that period. This problem was mitigated in the \citet{dong-ogle71} analysis by the
constraints they placed on the lens brightness and a crude constraint they placed on the lens-source
relative proper motion from the color-dependent centroid shift effect \citep{bennett06}.

\subsection{Comparison to Previous Analysis}
\label{sec-comp_prev}

Our conclusions are largely in line with those of \citet{dong-ogle71}, except that we do not have to 
make the assumption that the unlensed light coincident with the source in the 2005 and 2006
HST images is not due to a source companion. While our Keck data are consistent with the
existence of a faint source companion, our measurement of the $K$-band lens brightness indicates
excess $I$ and $V$-band flux seen in the HST images is primarily due to the lens star. 

The one \citet{dong-ogle71} claim that clearly seems to be incorrect is the claim that the
OGLE-2005-BLG-071L planetary host star has thick disk kinematics. \citet{dong-ogle71}  do
not provide many details about how they reach this conclusion, but a primary difference
appears to be the fact that they compare their estimated lens velocity to the local standard
of rest rather than the kinematics of the Galaxy at location of the lens, which was not well understood
in 2009 \citep{gaia18}.

It is also of interest to consider the accuracy of their measurement of the color dependent 
centroid shift measurement \citep{bennett06}, since this is of interest for WFIRST
\citep{aparna18}. \citet{dong-ogle71} measure offsets between the target (source plus lens blend)
centroid in the $I$ and $V$-band magnitudes of 
$\Delta r_{I-V,{\rm E}} = -0.52 \pm 0.20\,$mas and $\Delta r_{I-V,{\rm N}} = 0.22 \pm 0.20\,$mas
in the East and North directions, respectively. However, this includes both proper motion and 
parallax, so to compare to our measurements, we must convert this measurement to the
heliocentric reference frame. With our estimate of the lens system distance, $D_L$, the 
centroid shift becomes
$\Delta r_{I-V,{\rm E,H}} = -0.45 \pm 0.20\,$mas and $\Delta r_{I-V,{\rm N,H}} = 0.22 \pm 0.20\,$mas.
We can compare this to the value determined from our precise Keck $\bm{\mu}_{\rm rel,H}$
measurements. We find
$\Delta r_{I-V,{\rm E,H}} = -0.115 \pm 0.003\,$mas and $\Delta r_{I-V,{\rm N,H}} = 0.100 \pm 0.003\,$mas,
so the HST estimates are off by 1.7$\sigma$ to the East and 0.5$\sigma$ to the North. This implies that
the \citet{dong-ogle71} error bars are consistent with our measurement and that they
are not likely to be underestimated by as much as a factor of 2.

\section{Discussion and Conclusions}
\label{sec-conclude}

Our Keck AO follow-up observations have identified the OGLE-2005-BLG-071L planetary host 
star through measurements of the lens $K$-band magnitude, and the lens-source relative 
proper motion, $\bm{\mu}_{\rm rel,H}$. These measurements allow the determination of the
lens mass and distance through multiple, redundant constraints. The $\bm{\mu}_{\rm rel,H}$ measurement
can be combined with the partial microlensing parallax measurement from the microlensing light 
curve to yield the lens system masses, and the $\bm{\mu}_{\rm rel,H}$ measurement can also be
combined with the lens $K$-band magnitude to yield the lens mass and distance, as well. These 
determinations are consistent with each other, and we combine them to give host and
planet masses of $M_{\rm host} = 0.426\pm 0.037 \msun$, and $m_p = 3.27\pm 0.32 M_{\rm Jup}$,
with a projected separation of $a_\perp = 3.38 \pm 0.52\,$AU at a distance of $D_L = 3.46 \pm 0.33\,$kpc.
The excess flux seen in 2005 and 2006 HST observations is also consistent with these conclusions.

This confirms that tentative conclusion of \citet{dong-ogle71} that this event is a somewhat unexpected
sample of a super-Jupiter mass planet orbiting an M-dwarf. Such planetary systems are thought to be
rare \citep{laughlin04}. Since the wide model is slightly favored, the preferred orbital period is
12 years or more, and this means that many exoplanet Doppler radial velocity surveys do not monitor
their stars long enough to detect such planets. However, those surveys that do monitor the
radial velocity of low-mass stars for long periods of time claim a low frequency of such planets \citep{johnson10},
although they have have not yet done a complete statistical analysis. The microlensing statistical analysis
of \citet{suzuki16,suzuki18} seems quite consistent with the preliminary radial velocity results since it 
includes two planets with mass ratios of $q\sim 7\times 10^{-3}$, when we would expect $\sim 700$ if there
was an average of one such planet orbiting every star. So, we might expect an occurrence rate for such planets
of 1/300 or 1/400. The other similar planet in the \citet{suzuki16} sample
is MOA-2008-BLG-379 \citep{suzuki14,suzuki14e}.

It is expected that the occurrence rate for super-Jupiter planets should be higher for planets that orbit
high metallicity host stars \citep{fischer05}. Because the host star, OGLE-2005-BLG-071L, can now be
resolved from the source star in $K$-band AO images, it should now be possible to determine the host star
metallicity with moderate resolution $K$-band spectra \citep{babs10} using instruments such as 
Keck/OSIRIS and VLT/MUSE. Both of these instruments employ AO feeds and OH suppression that
are probably required for spectra of sufficient S/N to measure metallicity.

Finally, we should mention that this event is part of our Keck Key Strategic Mission Support (KSMS) 
program that aims to determine masses and distances for the vast majority of stars in the 
\citep{suzuki16} sample, as well as the MOA 9-year microlensing event sample that is now under
analysis. One important aspect of this program is that it can test an assumption that is currently used to 
to estimate the host mass for most of the planets found by microlensing
\citep{ogle390,dong-moa400}. For this event, 
a sub-Saturn mass planet \citep{aparna18}, and a Uranus-mass planet
\citep{bennett15,batista15}, we find that the measured host star mass is close to 
the Bayesian prediction, which relies upon the assumption that stars of any mass are equally likely 
to host the planet with the given mass ratio. This differs from the case of two other targets from
our KSMS program: MOA-2007-BLG-400 (Bhattacharya et al., in preparation) and MOA-2013-BLG-220
\citep{van19}. The mass measurement for these two events find masses that 
are at the 94th and 93rd percentile Bayesian analysis prediction based on the equal planet hosting
probability assumption. These events also have very similar mass ratios of $q = 2.2\times 10^{-3}$ and
$q = 3.3\times 10^{-3}$. This suggests that high mass stars may be much more likely to host 
planets beyond the snow line with mass ratios in the $0.002 < q < 0.0035$ range, but our
OGLE-2005-BLG-071 result suggests that this trend may not hold at higher mass ratios. 
Clearly, more mass microlens exoplanet host star mass measurements are needed to confirm 
such speculation.

\acknowledgments 
The Keck Telescope observations and analysis were supported by a NASA Keck PI Data Award, administered by the NASA Exoplanet Science Institute. Data presented herein were obtained at the W. M. Keck Observatory from telescope time allocated to the National Aeronautics and Space Administration through the agency’s scientific partnership with the California Institute of Technology and the University of California. The Observatory was made possible by the generous financial support of the W. M. Keck Foundation. DPB and AB were also supported by NASA through grant NASA-80NSSC18K0274. 
Some of this research has made use of the NASA Exoplanet Archive, which is operated by the California Institute of Technology, under contract with the National Aeronautics and Space Administration under the Exoplanet Exploration Program.
This work was supported by the University of Tasmania through the UTASFoundation and the endowed Warren Chair in Astronomy and the ANR COLD-WORLDS (ANR-18-CE31-0002).
A.U. was supported by the OGLE project funded by National Science Centre, Poland with the grant MAESTRO 2014/14/A/ST9/00121.


\begin{thebibliography}{}
%\bibitem[Alard(1997)]{alard97} Alard, C.\ 1997, \aap, 321, 424 
%\bibitem[Alard \& Lupton (1998)]{ala98}Alard, C. \& Lupton, R.H.\ 1998, \apj, 503, 325
%\bibitem[Albrow et al.(2000)]{albrow-97blg41} Albrow, M.D.\ 2000, \apj, 534, 894
%\bibitem[Albrow et al.(2009)]{albrow09} Albrow, M.~D., Horne, K., Bramich, D.~M., et al.\ 2009, \mnras, 397, 2099
%\bibitem[Alcock et al.(1995)]{macho-par1}Alcock, C., Allsman, R.~A., Alves, D., et al.~1995, \apjl, 454, L125
%\bibitem[Alcock et al.(1997)]{macho-95b30} Alcock, C., Allen, W.H., Allsman, R.A., et al.~1997, \apj, 491, 436
%\bibitem[An et al.(2007)]{an07} An, D., Terndrup, D.~M., Pinsonneault, M.~H., et al.\ 2007, \apj, 655, 233 
\bibitem[Anderson \& King (2000)]{andking00} Anderson, J.~\& King, I.~R.\ 2000, \pasp, 112, 1360
\bibitem[Anderson \& King (2004)]{andking04} Anderson, J.~\& King, I.~R.\ 2004, Hubble Space Telescope
   Advanced Camera for Surveys Instrument Science Report 04-15
\bibitem[Anderson \& King (2006)]{andking06} Anderson, J.~\& King, I.~R.,\ 2006, Hubble Space Telescope
   Advanced Camera for Surveys Instrument Science Report 2006-1
\bibitem[Anglada-Escud{\'e} et al.(2012)]{anglada-escude12_gj317} Anglada-Escud{\'e}, G., Boss, A.~P., Weinberger, A.~J., et al.\ 2012, The Astrophysical Journal, 746, 37
%\bibitem[Bachelet et al.(2012)]{bachelet12} Bachelet, E., Fouqu{\'e}, P., Han, C., et al.\ 2012, \aap, 547, A55
%\bibitem[Baraffe et al.(2005)]{baraffe05} Baraffe, I., et al.\ 2005, \aap, 436, L47
%\bibitem[Barry(2010)]{barry_psf} Barry, R.K., et al.\ 2010,  Proc.\ SPIE, 7731, 77313
%\bibitem[Batalha et al.(2013)]{kepler_16mon} Batalha, N.~M., Rowe, J.~F., Bryson, S.~T., et al.\ 2013, \apjs, 204, 24
\bibitem[Batista et al.(2011)]{batista11} Batista, V., Gould, A., Dieters, S., et al.\ 2011, \aap, 529, A102
\bibitem[Batista et al.(2014)]{batista14} Batista, V., Beaulieu, J.-P., Gould, A., et al.\ 2014, \apj, 780, 54 
\bibitem[Batista et al.(2015)]{batista15} Batista, V., Beaulieu, J.-P., Bennett, D.P., et al.\ 2015, \apj, 808, 170
\bibitem[Beaulieu et al.(2018)]{beaulieu18} Beaulieu, J.-P., Batista, V., Bennett, D.~P., et al.\ 2018, \aj, 155, 78
\bibitem[Beaulieu et al.(2006)]{ogle390} Beaulieu, J.-P., Bennett, D.~P., Fouqu{\'e}, P., et al.\ 2006, \nat, 439, 437
\bibitem[Beaulieu et al.(2016)]{beaulieu16} Beaulieu, J.-P., Bennett, D.~P., Batista, V., et al.\ 2016, \apj, 824, 83 
%\bibitem[Beichman et al.(2013)]{beichman13} Beichman, C., Gelino, C.~R., Kirkpatrick, J.~D., et al.\ 2013, \apj, 764, 101 
%\bibitem[Benedict et al.(2016)]{benedict16} Benedict, G.~F., Henry, T.~J., Franz, O.~G., et al.\ 2016, \aj, 152, 141
\bibitem[Bennett(2008)]{bennett_rev} Bennett, D.P, 2008, in Exoplanets, 
   Edited by John Mason.~Berlin: Springer.~ ISBN: 978-3-540-74007-0,  (arXiv:0902.1761)
\bibitem[Bennett(2010)]{bennett-himag} Bennett, D.P.\ 2010, \apj, 716, 1408
%\bibitem[Bennett et al.(1993)]{bennett-sod} Bennett, D.~P., Alcock, C., Allsman, R., et al.\ 1993, Bulletin of the American Astronomical Society, 25, 1402
%\bibitem[Bennett et al.(2018a)]{bennett18_wfirst} Bennett, D.~P., Akeson, R., Anderson, J., et al.\ 2018a, (arXiv:1803.08564)
%\bibitem[Bennett et al.(2010a)]{bennett_MPF} Bennett, D.~P., Anderson, J., Beaulieu, J.-P., et al.\ 2010a, arXiv:1012.4486
\bibitem[Bennett et al.(2006)]{bennett06} Bennett, D.~P., Anderson, J., Bond, I.~A., Udalski, A., \& Gould, A.\ 2006, \apjl, 647, L171
\bibitem[Bennett et al.(2007)]{bennett07} Bennett, D.P., Anderson, J., \& Gaudi, B.S.\ 2007, \apj, 660, 781
\bibitem[Bennett et al.(2014)]{bennett14} Bennett, D.~P., Batista, V., Bond, I.~A., et~al.\ 2014, \apj, 785, 155
\bibitem[Bennett et al.(2015)]{bennett15} Bennett, D.~P., Bhattacharya, A., Anderson, J., et al.\ 2015, \apj, 808, 169
%\bibitem[Bennett et al.(2008)]{bennett08}Bennett, D.~P., Bond, I.~A., Udalski, A., et al.\ 2008, \apj, 684, 663
\bibitem[Bennett et al.(2017)]{bennett_ob150954} Bennett, D.~P., Bond, I.~A., Abe, F., et al.\ 2017, \aj, 154, 68
%\bibitem[Bennett \& Rhie(1996)]{bennett96}Bennett, D.P. \& Rhie, S.H.\ 1996, \apj, 472, 660
%\bibitem[Bennett \& Rhie(2002)]{bennett02}Bennett, D.P. \& Rhie, S.H.\ 2002, \apj, 574, 985
\bibitem[Bennett et al.(2010b)]{bennett-ogle109} Bennett, D.~P., Rhie, S.~H., Nikolaev, S., et~al.\ 2010b, \apj, 713, 837
%\bibitem[Bennett et al.(2016)]{bennett16}Bennett, D.P., Rhie, S.H., Udalski, A., et al.\ 2016, \aj, 152, 125
%\bibitem[Bennett et al.(2012)]{bennett12} Bennett, D.~P., Sumi, T., Bond, I.~A., et al.\ 2012, \apj, 757, 119 
\bibitem[Bennett et al.(2018b)]{bennett_moa291} Bennett, D.~P., Udalski, A., Bond, I.~A., et al.\ 2018b, \aj, 156, 113
%\bibitem[Bennett et al.(2018a)]{bennett18} Bennett, D.~P., Udalski, A., Han, C., et al.\ 2018a, \aj, 155, 141 
%\bibitem[Bensby et al.(2011)]{bensby11} Bensby, T., Ad{\'e}n, D., Mel{\'e}ndez, J., et al.\ 2011, \aap, 533, A134
\bibitem[Bertin \& Arnouts(1996)]{sextractor} Bertin, E., $\&$ Arnouts, S., 1996, A$\&$AS, 117, 393
\bibitem[Bertin(2010a)]{Scamp2010} Bertin, E., 2010a, SCAMP: Automatic Astrometric and Photometric Calibration, ascl:1010.063
\bibitem[Bertin(2010b)]{Swarp2010} Bertin, E., 2010b, SWarp: Resampling and Co-adding FITS Images Together, ascl:1010.068
\bibitem[Bertin et al.(2012)]{Astromatic2012} Bertin, E., Delorme, P., and Bouy, H, 2012, Astrophysics and Space Science Proceedings, Vol. 29, 71
\bibitem[Bertin et al.(2002)]{SWarp} Bertin, E., Mellier, Y. , Radovich, M.,et al., 2002, The TERAPIX Pipeline, ASP Conference Series, Vol. 281, 228
%\bibitem[Bhattacharya et al.(2017)]{aparna17} Bhattacharya, A., Bennett, D.~P., Anderson, J., et al.\ 2017, \aj, 154, 59 
\bibitem[Bhattacharya et al.(2018)]{aparna18} Bhattacharya, A., Beaulieu, J.-P., Bennett, D.~P., et al.\ 2018, \aj, 156, 289 
%\bibitem[Bond et al.(2001)]{bond01} Bond, I.~A., Abe, F., Dodd, R.~J., et al.\ 2001, \mnras, 327, 868
%\bibitem[Bond et~al.(2004)]{bond04} Bond, I.~A., Udalski, A., Jaroszy{\'n}ski, M.\ 2004,  \apjl, 606, L155
%\bibitem[Bonfils et al.(2011)]{bonfils11} Bonfils, X., Delfosse, X., Udry, S., et al.\ 2011, arXiv:1111.5019
%\bibitem[Borucki et al.(2011)]{borucki11} Borucki, W.~J., Koch, D.~G., Basri, G., et al.\ 2011, \apj, 736, 19
%\bibitem[Boss(1997)]{boss97} Boss, A.~P.\ 1997, Science, 276, 1836
%\bibitem[Boss(2006)]{boss06} Boss, A.~P.\ 2006, \apj, 643, 501
%\bibitem[Boss(2006)]{boss06} Boss, A.P.\ 2006, \apjl, 644, L79
%\bibitem[Bowler et al.(2011)]{bowler11} Bowler, B.~P., Liu, M.~C., Kraus, A.~L., Mann, A.~W., \& 
%Ireland, M.~J.\ 2011, \apj, 743, 148
%\bibitem[Boyajian et al.(2014)]{boyajian14} Boyajian, T.S., van Belle, G., \& von Braun, K.,\  2014, \aj, 147, 47
%\bibitem[Bressan et al.(2012)]{bressan12_PARSEC} Bressan, A., Marigo, P., Girardi, L., et al.\ 2012, \mnras, 427, 127 
%\bibitem[Brown et al.(2013)]{lcogt} Brown, T.~M., Baliber, N., Bianco, F.~B., et al.\ 2013, \pasp, in press (arXiv:1305.2437) 
%\bibitem[Butler et al.(2006)]{butler-catalog} Butler, R.~P., Wright, J.~T., Marcy, G.~W., et al.\ 2006, \apj, 646, 505
%\bibitem[Burke et al.(2015)]{burke15} Burke, C.~J., Christiansen, J.~L., Mullally, F., et al.\ 2015, \apj, 809, 8 
%\bibitem[Bramich(2008)]{bramich08} Bramich, D.M.\ 2008, \mnras, 386, L77
%\bibitem[Cardelli et al.(1989)]{cardelli89}  Cardelli, J.A., Clayton, G.C., \& Mathis, J.S.\ 1989, \apj, 345, 245
%\bibitem[Cassan et al.(2012)]{cassan12}Cassan, A., Kubas, D., Beaulieu, J.-P., et al.\ 2012,  \nat, 481, 167
\bibitem[Carpenter(2001)]{2mass_cal} Carpenter, J.M.\ 2001, \aj 121, 2851
%\bibitem[Claret(2000)]{claret00} Claret, A.\ 2000, \aap, 363, 1081
%\bibitem[Chang \& Refsdal(1979)]{chang-refsdal79} Chang, K., \& Refsdal, S.\ 1979, \nat, 282, 561
%\bibitem[Chang \& Refsdal(1984)]{chang-refsdal84} Chang, K., \& Refsdal, S.\ 1979, \aap, 132, 168
%\bibitem[Chen et al.(2015)]{chen15_PARSEC} Chen, Y., Bressan, A., Girardi, L., et al.\ 2015, \mnras, 452, 1068 
%\bibitem[Chen et al.(2014)]{chen14_PARSEC} Chen, Y., Girardi, L., Bressan, A., et al.\ 2014, \mnras, 444, 2525 
%\bibitem[Christiansen et al.(2011)]{epoxi} Christiansen, J.L., et al.\ 2011, \apj, 726, 94
%\bibitem[Clarkson et al.(2008)]{clarkson08} Clarkson, W., Sahu, K., Anderson, J., et al.\ 2008, \apj, 684, 1110-1142 
%\bibitem[Cumming et al.(2008)]{cumming08}Cumming, A., Butler, R.~P., Marcy, G.~W., Vogt, S.~S., Wright, J.~T., \& Fischer, D.~A.\ 2008, \pasp, 120, 531
%\bibitem[D'Angelo et al.(2010)]{dangelo_book} D'Angelo, G., Durisen, R.~H., \& Lissauer, J.~J.\ 2010,
%   in Exoplanets, ed.\ S.\ Seager (Tucson, AZ: Univ. Arizona Press), 319
\bibitem[Delfosse et al.(2000)]{delfosse00} Delfosse, X., Forveille, T., S{\'e}gransan, D., et al.\ 2000, \aap, 364, 217
%\bibitem[Delorme et al.(2012)]{delhome12} Delorme, P., Gagn{\'e}, J., Malo, L., et al.\ 2012, \aap, 548, A26 
%\bibitem[Di Stefano \& Esin(1995)]{distefano95} Di Stefano, R., \& Esin, A.~A.\ 1995, \apjl, 448, L1
%\bibitem[Di Stefano \& Scalzo(1999)]{distefano99} Di Stefano, R., \& Scalzo, R.~A.\ 1999, \apj, 512, 579
\bibitem[Dolphin(2000)]{hstphot} Dolphin, A.~E.\ 2000, \pasp, 112, 1383 
%\bibitem[Dong et al.(2006)]{dong06} Dong, S., et al.\ 2006, , \apj, 642, 842
%\bibitem[Dong et al.(2007)]{dong-ogle05smc1} Dong, S., et al.\ 2007, \apj, 664, 862
\bibitem[Dong et al.(2009a)]{dong-moa400} Dong, S., Bond, I.~A., Gould, A., et al.\ 2009a, \apj, 698, 1826
\bibitem[Dong et al.(2009b)]{dong-ogle71} Dong, S., Gould, A., Udalski, A., et al.\ 2009b, \apj, 695, 970
%\bibitem[Doolin \& Blundell(2011)]{doolin11} Doolin, S., \& Blundell, K.~M.\ 2011, \mnras, 418, 2656
%\bibitem[Drimmel \& Spergel(2001)]{drimmel}  Drimmel, R., \& Spergel, D.~N.\ 2001, \apj, 556, 181
\bibitem[Fischer, \& Valenti(2005)]{fischer05} Fischer, D.~A., \& Valenti, J.\ 2005, \apj, 622, 1102
%\bibitem[Fruchter \& Hook(2002)]{drizzle} Fruchter A.S.\ 2002, \pasp, 114, 144
%\bibitem[Ford \& Rasio(2008)]{ford08} Ford, E.~B., \& Rasio, F.~A.\ 2008, \apj, 686, 621
%\bibitem[Fukui et al.(2015)]{fukui15} Fukui, A., Gould, A., Sumi, T., et al.\ 2015, \apj, 809, 74 
%\bibitem[Furusawa et al.(2013)]{moa328} Furusawa, K., Udalski, A., Sumi, T., et al.\ 2013, \apj, 779, 91
\bibitem[Gaia Collaboration et al.(2018a)]{gaia_dr2} Gaia Collaboration, Brown, A.~G.~A., Vallenari, A., et al.\ 2018a, \aap, 616, A1
\bibitem[Gaia Collaboration et al.(2018b)]{gaia18} Gaia Collaboration, Katz, D., Antoja, T., et al.\ 2018b, \aap, 616, A11
%\bibitem[Gaudi(2010)]{gaudi_rev}Gaudi, B.S.\ 2010, in Exoplanets, ed. S. Seager (Tucson: University of 
%Arizona Press), 79 (arXiv:1002.0332)
\bibitem[Gaudi(2012)]{gaudi_araa} Gaudi, B.~S.\ 2012, \araa, 50, 411
\bibitem[Gaudi et al.(2008)]{gaudi-ogle109} Gaudi, B.~S., Bennett, D.~P., Udalski, A., et al.\ 2008, Science, 319, 927
%\bibitem[Gaudi \& Gould(1997)]{gaudi97} Gaudi, B.S., \& Gould, A.\ 1997, \apj, 486, 85
\bibitem[Ghosh et al.(2004)]{ghosh04} Ghosh, H., DePoy, D.~L., Gal-Yam, A., et al.\ 2004, \apj, 615, 450
\bibitem[Gonzalez et al.(2011)]{vvv_extinct} Gonzalez, O.~A., Rejkuba, M., Zoccali, M., Valenti, E., \& Minniti, D.\ 2011, \aap, 534, A3
%\bibitem[Gould(1992)]{gould-par1} Gould, A.\ 1992, \apj, 392, 442
%\bibitem[Gould(2004)]{gould-jerk} Gould, A.\ 2004, \apj, 606, 319
%\bibitem[Gould(2008)]{gould-hex} Gould, A.\ 2008, \apj, 681, 1593
\bibitem[Gould(2014)]{gould-1dpar} Gould, A.\ 2014, J.\ Kor.\ Ast.\ Soc., 47, 215
%\bibitem[Gould et al.(2010)]{gould_col} Gould, A., Dong, S., Bennett, D.~P., et al.\ 2010a, \apj, 710, 1800
%\bibitem[Gould et al.(2010b)]{gould10}Gould, A., Dong, S., Gaudi, B.S., et al.\ 2010b,  \apj, 720, 1073
%\bibitem[Gould et al.(2004)]{gould04}Gould, A., Gaudi, B.S., \& Han, C., 2004, arXiv:astro-ph/0405217
%\bibitem[Gould \& Loeb(1992)]{gouldloeb92} Gould, A. \& Loeb, A. 1992, \apj, 396, 104
\bibitem[Gould et al.(1994)]{gould-diskorhalo}  Gould, A., Miralda-Escude, J., \& Bahcall, J.~N.\ 1994, \apjl, 423, L105
%\bibitem[Gould et al.(2006)]{gould06} Gould, A., Udalski, A., An, D., et al.\ 2006, \apjl, 644, L37
%\bibitem[Gould et al.(2009)]{gould09} Gould, A., Udalski, A., Monard, B., et al.\ 2009, \apjl, 698, L147
%\bibitem[Gould et al.(2014)]{gould14} Gould, A., Udalski, A., Shin, I.-G., et al.\ 2014, Science, 345, 46 
%\bibitem[Gould \& Yee(2013)]{gould_yee_terpar13} Gould, A., \& Yee, J.~C.\ 2013, \apj, 764, 107
%\bibitem[Green et al.(2012)]{WFIRST_rep} Green, J., Schechter, P., Baltay, C., et al.\ 2012, arXiv:1208.4012 
%\bibitem[Griest \& Hu(1992)]{griest92} Griest, K., \& Hu, W.\ 1992, \apj, 397, 362
%\bibitem[Griest \& Safizadeh(1998)]{griest98} Griest, K., \& Safizadeh, N.\ 1998, \apj, 500, 37
\bibitem[Gubler $\&$ Tytler(1998)]{gubler} Gubler, J., $\&$ Tytler, D., 1998, PASP, 110, 738
%\bibitem[Han et al.(2016)]{han_ob130723} Han, C., Bennett, D.~P., Udalski, A., \& Jung, Y.~K.\ 2016, \apj, in press (arXiv:1604.06533)
%\bibitem[Han \& Gould(1997)]{han97} Han , C., \& Gould, A.\ 1997, \apj, 480, 196
%\bibitem[Guillochon et al.(2011)]{guillochon11} Guillochon, J., Ramirez-Ruiz, E., \& Lin, D.\ 2011, \apj, 732, 74 
%\bibitem[Han et al.(2005)]{han_widepl2005} Han, C., Gaudi, B.~S., An, J.~H., \& Gould, A.\ 2005, \apj, 618, 962 
\bibitem[Han et al.(2019)]{han_ob181011} Han, C., Bennett, D.~P., Udalski, A., et al.\ 2019, The Astronomical Journal, 158, 114
%\bibitem[Han \& Kang(2003)]{han_widepl2003} Han, C., \& Kang, Y.~W.\ 2003, \apj, 596, 1320
%\bibitem[Han et al.(2017)]{han_ob160263} Han, C., Udalski, A., Gould, A., et al.\ 2017, \aj, 154, 133
%\bibitem[Hartman et al.(2004)]{hartmanISIS} Hartman, J. D., Bakos, G., Stanek, K. Z., \& Noyes, R. W. 2004, AJ, 128, 1761
%\bibitem[Henderson et al.(2014)]{henderson14} Henderson, C.~B., Park, H., Sumi, T., et al.\ 2014, \apj, 794, 71
\bibitem[Henry et al.(1999)]{henry99} Henry, T.~J., Franz, O.~G., Wasserman, L.~H., et al.\ 1999, \apj, 512, 864 
\bibitem[Henry \& McCarthy(1993)]{henry93} Henry, T.~J., \& McCarthy, D.~W., Jr.\ 1993, \aj, 106, 773
\bibitem[Hirao et al.(2017)]{hirao17} Hirao, Y., Udalski, A., Sumi, T., et al.\ 2017, \aj, 154, 1 
%\bibitem[Hilton(2011)]{hilton11} Hilton, E.~J.\ 2011, PhD Thesis, University of Washington
%\bibitem[Hilton et al.(2010)]{hilton10} Hilton, E.~J., Hawley, S.~L., Kowalski, A.~F., \& Holtzman, J.\ 2010, arXiv:1012.0577 
%\bibitem[Holman \& Wiegert(1999)]{holman99} Holman, M.~J., \& Wiegert, P.~A.\ 1999, \aj, 117, 621 
%\bibitem[Heyrovsky(2003)]{heyrovsky03} Heyrovsk\'y, D.\ 2003, \apj, 594, 464
%\bibitem[Heyrovsky(2007)]{heyrovsky07} Heyrovsk\'y, D.\ 2007, \apj, 656, 483
%\bibitem[Holtzman et al.(1998)]{holtzman98} Holtzman, J.~A., Watson, A.~M., Baum, W.~A., et al.\ 1998, \aj, 115, 1946 
%\bibitem[Howard et al.(2010)]{howard10}Howard, A.W. et al.\ 2010, Science, 330, 653
%\bibitem[Hubickyj et al.(2005)]{hubickyj05} Hubickyj, O., Bodenheimer, P., \& Lissauer, J.J.\ 2005, Icarus, 179, 415
%\bibitem[Hwang et al.(2018)]{hwang18} Hwang, K.-H., Udalski, A., Shvartzvald, Y., et al.\ 2018, \aj, 155, 20
%\bibitem[Hwang et al.(2019)]{hwang_kb161107} Hwang, K.-H., Ryu, Y.-H., Kim, H.-W., et al.\ 2019, \aj,, 157, 23
%\bibitem[Ida \& Lin(2004)]{idalin04} Ida, S.\ \& Lin, D.N.C.\ 2004, \apj, 604, 388
%\bibitem[Ida \& Lin(2005)]{ida05} Ida, S., \& Lin, D.N.C.\ 2005, \apj, 626, 1045
%\bibitem[Janczak et al.(2010)]{janczak10} Janczak, J., Fukui, A., Dong, S., et al.\ 2010, \apj, 711, 731
\bibitem[Johnson et al.(2007)]{johnson07} Johnson, J.~A., Butler, R.~P., Marcy, G.~W., et al.\ 2007, \apj, 670, 833
\bibitem[Johnson et al.(2010)]{johnson10} Johnson, J.~A., Aller, K.~M., Howard, A.~W., \& Crepp, J.~R.\ 2010, \pasp, 122, 905 
\bibitem[Jung et al.(2015)]{jung_ob130102} Jung, Y.~K., Udalski, A., Sumi, T., et al.\ 2015, \apj, 798, 123
%\bibitem[Kaib et al.(2013)]{kaib13} Kaib, N.~A., Raymond, S.~N., \& Duncan, M.\ 2013, \nat, 493, 381 
\bibitem[Kennedy \& Kenyon(2008)]{kennedy_snowline} Kennedy, G.~M., \& Kenyon, S.~J.\ 2008, \apj, 673, 502 
%\bibitem[Kennedy et al.(2006)]{kennedy-searth} Kennedy, G.M., Kenyon, S.J.,  \& Bromley, B.C.\ 2006, \apjl 650, L139
\bibitem[Kenyon \& Hartmann(1995)]{kenyon95} Kenyon, S.~J., \& Hartmann, L.\ 1995, \apjs, 101, 117 
%\bibitem[Kervella et al.(2004)]{kervella_dwarf} Kervella, P., Th{\'e}venin, F., Di Folco, E., \& S{\'e}gransan, D.\ 2004, \aap, 426, 297
%\bibitem[Kervella et al.(2004)]{kervella04g} Kervella P., et al.\ 2004,  \aap, 428, 587
%\bibitem[Kervella \& Fouqu{\'e}(2008)]{kervella08} Kervella, P., \& Fouqu{\'e}, P.\ 2008, \aap, 491, 855
%\bibitem[Kim et al.(2013)]{kmtnet_pipe} Kim, D.~J., Lee, C.~U., Kim, S.~L., \& Park, B.~G.\ 2013, Pub.\ Korean Astron.\ Soc., 28, 1
%\bibitem[Kim et al.(2010)]{kmtnet10} Kim, S.-L., Park, B.-G., Lee, C.-U., et al.\ 2010, \procspie, 7733, 77733
%\bibitem[Kim et al.(2016)]{kmtnet} Kim, S.-L., Lee, C.-U., Park, B.-G., et al.\ 2016, Journal of Korean Astronomical Society, 49, 37 
\bibitem[King(1983)]{king1983} King, I.~R., 1983, PASP, 95, 163
%\bibitem[Koshimoto et al.(2017a)]{kosh17_ob120950} Koshimoto, N., Udalski, A., Beaulieu, J.~P., et al.\ 2017a, \aj, 153, 1 
\bibitem[Koshimoto et al.(2017b)]{kosh17_mb16227} Koshimoto, N., Shvartzvald, Y., Bennett, D.~P., et al.\ 2017b, \aj, 154, 3
\bibitem[Koshimoto et al.(2014)]{koshimoto14} Koshimoto, N., Udalski, A., Sumi, T., et al.\ 2014, \apj, 788, 128
%\bibitem[Kowalski et al.(2009)]{kowalski09} Kowalski, A.~F., Hawley, S.~L., Hilton, E.~J., et al.\ 2009, \aj, 138, 633
%\bibitem[Kowalski et al.(2010)]{kowalski10} Kowalski, A.~F., Hawley, S.~L., Holtzman, J.~A., Wisniewski, J.~P., \& Hilton, E.~J.\ 2010, \apjl, 714, L98
%\bibitem[Koz{\l}owski et al.(2006)]{koz06} Koz{\l}owski, S., Wo{\'z}niak, P.~R., Mao, S., et al.\ 2006, \mnras, 370, 435
%\bibitem[Kubas et al.(2012)]{moa192_naco} Kubas, D., Beaulieu, J.~P., Bennett, D.~P., et al.\ 2012, \aap, 540, A78
%\bibitem[Kurucz(1993a)]{kurucz93a} Kurucz, R.L.\ 1993a, Kurucz CD-ROM 16, (SAO, Cambridge, MA, 1993).
%\bibitem[Kurucz(1993b)]{kurucz93b} Kurucz, R.L.\ 1993b, Kurucz CD-ROM 17,  (SAO, Cambridge, MA, 1993)
%\bibitem[Kurucz(1994)]{kurucz94} Kurucz, R.L.\ 1994, Kurucz CD-ROM 19,  (SAO, Cambridge, MA, 1993)
%\bibitem[Kurucz(1996)]{kurucz} Kurucz, R.L.\ 1996, ASP Conference Series, 108, 2
\bibitem[Laughlin et al.(2004)]{laughlin04} Laughlin, G.  Bodenheimer, P.\ \& Adams, F.C.\ 2004, \apjl, 612, L73
%\bibitem[Lecar(2006)]{lecar_snowline} Lecar, M., Podolak, M., Sasselov, D., \& Chiang, E.\ 2006, \apj, 640, 1115
%\bibitem[Levison et~al.(1998)]{levison98} Levison, H.~F., Lissauer, J.~J., Duncan, M.~J.\ 1998, \aj, 116, 1998
\bibitem[Lindegren(2018)]{lindegren18} Lindegren, L.\ 2018, Gaia report GAIA-C3-TN-LU-LL-124-01
%\bibitem[Lissauer(1993)]{lissauer_araa} Lissauer, J.J.\ 1993, Ann.\ Rev.\ Astron.\ Ast., 31, 129
%\bibitem[Lo Curto et al.(2010)]{locurto10} Lo Curto, G., Mayor, M., Benz, W., et al.\ 2010, Astronomy and Astrophysics, 512, A48
%\bibitem[Malmberg et al.(2011)]{malmberg11} Malmberg, D., Davies, M.~B., \& Heggie, D.~C.\ 2011, \mnras, 411, 859
%\bibitem[Lu et al.(2014)]{lu14} Lu, J.~R., Neichel, B., Anderson, J., et al.\ 2014, \procspie, 9148, 91480B 
%\bibitem[Lu et al.(2016)]{lu16} Lu, J.~R., Sinukoff, E., Ofek, E.~O., Udalski, A., \& Kozlowski, S.\ 2016, \apj, 830, 41 
%\bibitem[Mao \& Paczy\'{n}ski(1991)]{mao91}Mao, S., \& Paczy\'{n}ski, B.\ 1991, \apj, 374, L37
%\bibitem[Mayor \& Queloz(2012)]{mayor12} Mayor, M., \& Queloz, D.\ 2012, \nar, 56, 19
\bibitem[Minniti et al.(2010)]{minniti-vvv} Minniti, D., Lucas, P.~W., Emerson, J.~P., et al.\ 2010, \na, 15, 433 
%\bibitem[Musielak et al.(2005)]{musielak05} Musielak, Z.~E., Cuntz, M., Marshall, E.~A., \& Stuit, T.~D.\ 2005, \aap, 434, 355 
%\bibitem[Miyake et al.(2011)]{miyake11} Miyake, N., Sumi, T., Dong, S., et al.\ 2011, \apj, 728, 120
%\bibitem[Montet et al.(2013)]{montet13} Montet, B.T., Crepp, J.R., Johnson, J.A., Howard, A.W., \& Marcy, G.W.\ 2013, \apj, submitted (arXiv:1307.5849 )
%\bibitem[Movshovitz \& Podolak(2008)]{movshovitz08} Movshovitz, N., \& Podolak, M.\ 2008, Icarus, 194, 368
%\bibitem[Mr{\'o}z et al.(2017a)]{mroz17_ob160596} Mr{\'o}z, P., Han, C., Udalski, A., et al.\ 2017a, \aj, 153, 143 
%\bibitem[Mr{\'o}z et al.(2017b)]{mroz17_2pl} Mr{\'o}z, P., Udalski, A., Bond, I.~A., et al.\ 2017b, \aj, 154, 205 
%\bibitem[Mullally et al.(2016)]{mullaly16} Mullally, F., Coughlin, J.~L., Thompson, S.~E., et al.\ 2016, arXiv:1602.03204 
\bibitem[Muraki et al.(2011)]{muraki11} Muraki, Y., Han, C., Bennett, D.~P., et al.\ 2011, \apj, 741, 22
%\bibitem[Nagasawa \& Ida(2011)]{nagasawa11} Nagasawa, M., \& Ida, S.\ 2011, \apj, 742, 72 
\bibitem[Nataf et al.(2013)]{nataf13} Nataf, D.~M., Gould, A., Fouqu{\'e}, P., et al.\ 2013, \apj, 769, 88 
\bibitem[Nishiyama et al.(2009)]{nish09} Nishiyama, S., Tamura, M., Hatano, H., et al.\ 2009, \apj, 696, 1407
%\bibitem[Park et al.(2012)]{kmtnet} Park, B.-G., Kim, S.-L., Lee, J.~W., et al.\ 2012, \procspie, 8444, 844447-1
%\bibitem[Pejcha \& Heyrovsky(2009)]{pei_hey} Pejcha, O., \& Heyrovsk\'y, D.\ 2009, \apj, 690, 1772
%\bibitem[Penny et al.(2013)]{penny13} Penny, M.~T., Kerins, E., Rattenbury, N., et al.\ 2013,  \mnras, submitted (arXiv:1206.5296)
%\bibitem[Penny et al.(2018)]{penny18} Penny, M.~T., Gaudi, B.~S., Kerins, E., et al.\ 2018, arXiv:1808.02490
%\bibitem[Pietrukowicz et al.(2011)]{m22ml} Pietrukowicz, P., Minniti, D., Jetzer, P., Alonso-Garcia, J., \& Udalski, A.\ 2011, \apj, 744, L18
%\bibitem[Poindexter et al.(2005)]{poindexter05} Poindexter, S., et al.\ 2005, \apj, 633, 914
\bibitem[Poleski et al.(2017)]{poleski_mb12006} Poleski, R., Udalski, A., Bond, I.~A., et al.\ 2017, \aap, 604, A103
\bibitem[Poleski et al.(2014)]{poleski_ob120406} Poleski, R., Udalski, A., Dong, S., et al.\ 2014, The Astrophysical Journal, 782, 47
%\bibitem[Pollack et al.(1996)]{pollack96} Pollack, J.~B., Hubickyj, O., Bodenheimer, P., et al.\ 1996, \icarus, 124, 62
%\bibitem[Popowski et al.(2003)]{pop_extinct} Popowski, P., Cook, K.~H., \& Becker, A.~C.\ 2003, \aj, 126, 2910
%\bibitem[Quanz et al.(2012)]{quanz12} Quanz, S.~P., Lafreni{\`e}re, D., Meyer, M.~R., Reggiani, M.~M., \& Buenzli, E.\ 2012, \aap, 541, A133 
%\bibitem[Ranc et al.(2015)]{ranc15} Ranc, C., Cassan, A., Albrow, M.~D., et al.\ 2015, \aap, 580, A125
%\bibitem[Ranc et al.(2018)]{ranc18} Ranc, C., et al.\ 2018, in preparation
%\bibitem[Rafikov(2011)]{rafikov11} Rafikov, R.\ 2011, \apj, 727, 86
%\bibitem[Rattenbury et al.(2007)]{rattenbury07} Rattenbury, N.~J., Mao, S., Debattista, V.~P., et al.\ 2007, \mnras, 378, 1165
%\bibitem[Rattenbury et al.(2015)]{rattenbury15} Rattenbury, N.~J., Bennett, D.~P., Sumi, T., et al.\ 2015, \mnras, 454, 946 
%\bibitem[Rattenbury et al.(2017)]{rattenbury17} Rattenbury, N.~J., Bennett, D.~P., Sumi, T., et al.\ 2017, \mnras, 466, 2710 
%\bibitem[Refsdal(1966)]{refsdal-par} Refsdal, S. 1966, \mnras, 134, 315
\bibitem[Rhie et al.(1999)]{rhie_98smc1} Rhie, S.~H., Becker, A.~C., Bennett, D.~P., et al.\ 1999, \apj, 522, 1037
%\bibitem[Rhie et al.(2000)]{rhie00} Rhie, S.~H., Bennett, D.~P., Becker, A.~C., et al.\ 2000, \apj, 533, 378
%\bibitem[Robin et al.(2003)]{robin03} Robin, A.~C., Reyl{\'e}, C., Derri{\`e}re, S., \& Picaud, S.\ 2003, \aap, 409, 523
\bibitem[Rojas-Ayala et al.(2010)]{babs10} Rojas-Ayala, B., Covey, K.~R., Muirhead, P.~S., and Lloyd, J.~P.\ 2010, \apjl, 720, L113
%\bibitem[Sako et al.(2008)]{sako_moacam3} Sako, T., Sekiguchi, T., Sasaki, M., et al.\ 2008, Experimental Astronomy, 22, 51
%\bibitem[Schechter, Mateo, \& Saha(1993)]{dophot} Schechter, P.~L., Mateo, M., \& Saha, A.\ 1993, \pasp, 105, 1342
\bibitem[Service et al.(2016)]{distortion} Service, M., Lu, J.~R., Campbell, R., et al.\ 2016, \pasp, 128, 095004
\bibitem[Shin et al.(2016)]{shin16} Shin, I.-G., Ryu, Y.-H., Udalski, A., et al.\ 2016, Journal of Korean Astronomical Society, 49, 73
\bibitem[Shin et al.(2019)]{shin19_kb171038_1146} Shin, I.-G., Ryu, Y.-H., Yee, J.~C., et al.\ 2019, The Astronomical Journal, 157, 146
%\bibitem[Shvartzvald et al.(2018)]{shvartzvald18} Shvartzvald, Y., Calchi Novati, S., Gaudi, B.~S., et al.\ 2018, \apjl, 857, L8 
\bibitem[Shvartzvald et al.(2014)]{shvartzvald14} Shvartzvald, Y., Maoz, D., Kaspi, S., et al.\ 2014, \mnras, 439, 604
%\bibitem[Shvartzvald et al.(2016)]{shvartzvald16} Shvartzvald, Y., Maoz, D., Udalski, A., et al.\ 2016, \mnras, 457, 4089 
%\bibitem[Skowron(2011)]{skowron11} Skowron, J., et al.\ 2011, \apj, submitted, (arXiv:1101.3312)
%\bibitem[Skowron et al.(2013)]{skowron13} Skowron, J., et al.\ 2011, \apj, submitted
%\bibitem[Smith et al.(2002)]{smith02} Smith, M.C., Mao, S., \& Wo\'zniak, P.\ 2002, \mnras, 332, 962
%\bibitem[Smith et al.(2003)]{smith03} Smith, M.C., Mao, S., \& Paczy{\'n}ski, B.\ 2003, \mnras, 339, 925
%\bibitem[Spergel et al.(2015)]{WFIRST_AFTA} Spergel, D., Gehrels, N., Baltay, C., et al.\ 2015, arXiv:1503.03757 
%\bibitem[Stetson(1994)]{allframe} Stetson, P.B.\ 1994, \pasp, 106, 250
\bibitem[Street et al.(2013)]{street_mb10073} Street, R.~A., Choi, J.-Y., Tsapras, Y., et al.\ 2013, \apj, 763, 67
%\bibitem[Street et al.(2016)]{street16} Street, R.~A., Udalski, A., Calchi Novati, S., et al.\ 2016, \apj, 819, 93
\bibitem[Stetson(1987)]{Daophot} Stetson, P.~B., \ 1987, PASP, 99, 191S
%\bibitem[Stubbs et al.(2007)]{stubbs07} Stubbs, C.W., et al.\ 2007, \pasp, 119, 1163
%\bibitem[Sumi et al.(2010)]{sumi10}Sumi, T., Bennett, D.~P., Bond, I.~A. et al.\ 2010,  \apj, 710, 1641
%\bibitem[Sumi et al.(2011)]{sumi11}Sumi, T., Kamiya, K., Bennett, D.~P., et al.\ 2011,  \nat, 473, 349
%\bibitem[Sumi et al.(2016)]{sumi16} Sumi, T., Udalski, A., Bennett, D.~P., et al.\ 2016, \apj, 825, 112 
%\bibitem[Sumi et al.(2004)]{sumi04} Sumi, T., Wu, X., Udalski, A., et al.\ 2004, \mnras, 348, 1439
\bibitem[Suzuki et al.(2018)]{suzuki18} Suzuki, D., Bennett, D.~P., Ida, S., et al.\ 2018, \apjl, 869, L34
\bibitem[Suzuki et al.(2016)]{suzuki16} Suzuki, D., Bennett, D.~P., Sumi, T., et al.\ 2016, \apj, 833, 145
\bibitem[Suzuki et al.(2014)]{suzuki14} Suzuki, D., Udalski, A., Sumi, T., et al.\ 2014, \apj, 780, 123 
\bibitem[Suzuki et al.(2014e)]{suzuki14e} Suzuki, D., Udalski, A., Sumi, T., et al.\ 2014e, \apj, 788, 97 
\bibitem[Szyma{\'n}ski et al.(2011)]{ogle3-phot} Szyma{\'n}ski, M.~K., Udalski, A., Soszy{\'n}ski, I., et al.\ 2011, \actaa, 61, 83 
%\bibitem[Tang et al.(2014)]{tang14_PARSEC} Tang, J., Bressan, A., Rosenfield, P., et al.\ 2014, \mnras, 445, 4287 
%\bibitem[Thommes et al.(2008)]{thommes08} Thommes, E.W., Matsumura, S., \& Rasio F.A.\ 2008, Science, 321, 814
%\bibitem[Tomaney \& Crotts (1996)]{tom96}Tomaney, A.B. \& Crotts, A.P.S.\ 1996, \aj 112, 2872
%\bibitem[Traub(2011)]{traub11} Traub, W.\ 2011, ApJ, submitted.
\bibitem[Tsapras et al.(2014)]{tsapras_ob120406} Tsapras, Y., Choi, J.-Y., Street, R.~A., et al.\ 2014, \apj, 782, 48
%\bibitem[Twicken et al.(2016)]{kepler_q17} Twicken, J.~D., Jenkins, J.~M., Seader, S.~E., et al.\ 2016, ApJ, submitted (arXiv:1604.06140)
%\bibitem[Udalski(2003a)]{udalski-ext} Udalski, A.\ 2003a, \apj, 590, 284
%\bibitem[Udalski(2003)]{ogle-pipeline} Udalski, A.\ 2003, \actaa, 53, 291
\bibitem[Udalski et al.(2005)]{udalski05} Udalski, A., Jaroszy{\'n}ski, M., Paczy{\'n}ski, B., et al.\ 2005, \apjl, 628, L109
%\bibitem[Udalski et al.(2015b)]{udalski_ob130723} Udalski, A., Jung, Y.~K., Han, C., et al.\ 2015b, \apj, 812, 47 
%\bibitem[Udalski et al.(1994)]{ogle-ews} Udalski, A., Szyma\'{n}ski, M., Ka{\l}u\.{z}ny, J., Kubiak, M., Mateo, M., Krzmi\'{n}ski, W., \& \pac, B.\ 1994, Acta Astron., 44, 227
%\bibitem[Udalski et al.(2002)]{uda02} Udalski, A., Szymanski, M., Kubiak, M., et al.\ 2002, \actaa, 52, 217
%\bibitem[Udalski et al.(2008)]{udalski08} Udalski, A., et al.\ 2008, Acta Astron., 58, 69
%\bibitem[Udalski et al.(2015a)]{ogle4} Udalski, A., Szyma{\'n}ski, M.~K., \& Szyma{\'n}ski, G.\ 2015a, \actaa, 65, 1 
%\bibitem[Udalski et al.(2018)]{udalski18} Udalski, A., Ryu, Y.-H., Sajadian, S., et al.\ 2018, \actaa, 68, 1 
\bibitem[Vandorou et al.(2019)]{van19} Vandorou, A., Bennett, D.~P., Beaulieu, J.-P., et al.\ 2019, arXiv e-prints, arXiv:1909.04444
%\bibitem[Veras et al.(2011)]{veras11}  Veras, D., Wyatt, M.~C., Mustill, A.~J., Bonsor, A., \& Eldridge, J.~J.\ 2011, \mnras, 417, 2104
%\bibitem[Veras \& Raymond(2012)]{veras12} Veras, D., \& Raymond, S.~N.\ 2012, \mnras, 421, L117
%\bibitem[Veras \& Tout(2012)]{veras_tout12} Veras, D., \& Tout, C.~A.\ 2012, \mnras, 422, 1648 
%\bibitem[Voyatzis et al.(2013)]{voyatzis13} Voyatzis, G., Hadjidemetriou, J.~D., Veras, D., \& Varvoglis, H.\ 2013, \mnras, 430, 3383 
%\bibitem[Wambsganss(2011)]{wamb11} Wambsganss, J.\ 2011, \nat, 473, 289
%\bibitem[Ward(1997)]{ward97} Ward, W.R.\ 1997, Icarus, 126, 261
%\bibitem[Witt(1995)]{witt95} Witt, H.J.\ 1995, \apj, 449, 42
%\bibitem[Wright \& Gaudi(2013)]{wright_gaudi_book2013} Wright, J.~T., \& Gaudi, B.~S.\ 2013, Planets, Stars and Stellar Systems.~Volume 3: Solar and Stellar Planetary Systems, 489 
%\bibitem[Yee(2013)]{yee_WFIRST_par} Yee, J.~C.\ 2013, \apjl, 770, L31 
%\bibitem[Yee et al.(2012)]{yee12} Yee, J.~C., Shvartzvald, Y., Gal-Yam, A., et al.\ 2012, \apj, 755, 102
%\bibitem[Yee et al.(2009)]{yee09} Yee, J.~C., Udalski, A., Sumi, T., et al.\ 2009, \apj, 703, 2082 
\bibitem[Yelda et al.(2010)]{refraction} Yelda, S., Lu, J.~R., Ghez, A.~M., et al.\ 2010, \apj, 725, 331
%\bibitem[Yoo et al.(2004)]{yoo_rad} Yoo, J., DePoy, D.~L., Gal-Yam, A., et al.\ 2004, \apj, 603, 139
\bibitem[Zang et al.(2018)]{zang18_kb161397} Zang, W., Hwang, K.-H., Kim, H.-W., et al.\ 2018, \aj, 156, 236
\end{thebibliography}
\end{document}